\documentclass[11pt]{article}%
\usepackage{amsmath}
\usepackage{amsfonts}
\usepackage{amssymb}
\usepackage{graphicx}%
\ifx\pdfoutput\relax\let\pdfoutput=\undefined\fi
\newcount\msipdfoutput
\ifx\pdfoutput\undefined\else
\ifcase\pdfoutput\else
\ifx\paperwidth\undefined\else
\ifdim\paperheight=0pt\relax\else\pdfpageheight\paperheight\fi
\ifdim\paperwidth=0pt\relax\else\pdfpagewidth\paperwidth\fi
\fi\fi\fi
\begin{document}

\title{Propagation of gamma rays and production of free electrons in air}
\author{Y. S. Dimant, G. S. Nusinovich, P. Sprangle, J. Penano, C. A.
\and Romero-Talamas, \allowbreak and V. L. Granatstein}
\maketitle

\begin{abstract}
A new concept of remote detection of concealed radioactive materials has been
recently proposed \cite{Gr.Nusin.2010}-\cite{NusinSprangle}. It is based on
the breakdown in air at the focal point of a high-power beam of
electromagnetic waves produced by a THz gyrotron. To initiate the avalanche
breakdown, seed free electrons should be present in this focal region during
the electromagnetic pulse. This paper is devoted to the analysis of production
of free electrons by gamma rays leaking from radioactive materials. Within a
hundred meters from the radiation source, the fluctuating free electrons
appear with the rate that may exceed significantly the natural background
ionization rate. During the gyrotron pulse of about 10 microsecond length,
such electrons may seed the electric breakdown and create sufficiently dense
plasma at the focal region to be detected as an unambiguous effect of the
concealed radioactive material.

\end{abstract}

\section{Introduction}

A new concept of remote detection of concealed radioactive materials has been
recently proposed \cite{Gr.Nusin.2010}. This concept is based on breakdown in
free air at the focal point of a high-power beam of electromagnetic waves. To
realize such breakdown, the wave amplitude in a focal region should exceed the
breakdown threshold. To initiate the avalanche breakdown process, seed free
electrons should be present in this focal region during the electromagnetic
pulse. When the wavelength is short enough (sub-THz or THz frequency range)
the wave beam can be focused in a spot with dimensions on the order of a
wavelength and, then, the total volume where the wave amplitude exceeds the
breakdown threshold can be rather small. This fact allows one to realize
conditions \cite{Gr.Nusin.2010} when the breakdown rate in the case of the
ambient electron density is rather low. So in the cases when the breakdown
rate is much higher than the expected one can conclude that in the vicinity of
a focused wave beam there is a hidden source of radioactive material which
ionizes the air. Some issues important for realizing this concept are
discussed in Refs. \cite{NusiInfr} and \cite{NusinSprangle}.

\begin{figure}
[ptb]
\begin{center}
\includegraphics[
natheight=2.652100in,
natwidth=4.673500in,
height=2.6521in,
width=4.6735in
]%
{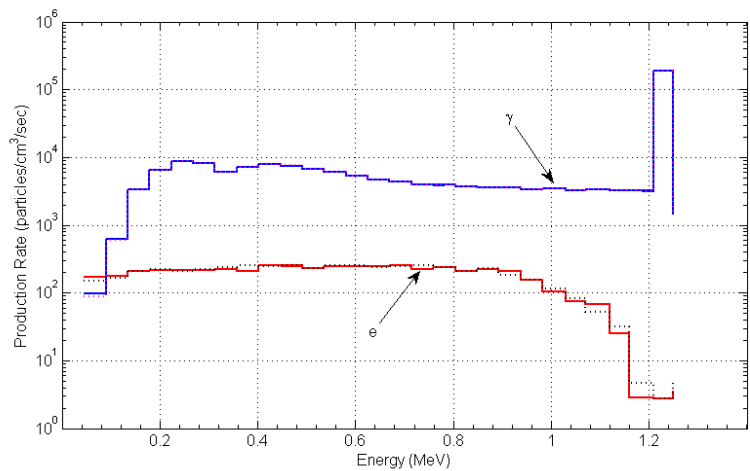}%
\caption{Results of Monte-Carlo simulations (MCNP code) for 1 Curie of
$_{27}^{60}$Co showing the energy spectrum of electrons (red) and photons
(blue) in a 1cm$^{3}$ tally volume outside a 0.5 cm thick steel shipping
container. Solid lines are the results of calculations for the case when
electron cutoff energy inside the steel wall adjacent to tally volume is set
to 1 keV, while dotted lines are for electron cutoff energy set to 100 keV. }%
\label{Fig.MCNP}%
\end{center}
\end{figure}

In the present paper, we analyze another issue important for realizing this
concept which was only briefly discussed in previous references, viz.
propagation of $\gamma$-rays in air and production of free electrons by these
quanta, mainly due to inelastic Compton scattering. It should be noted that,
in principle, due to the atomic reaction radioactive materials emit both
MeV-scale $\gamma$-rays and $\beta$-electrons. For example, a disintegrating
atom of $_{27}^{60}$Co produces one $2.505$~MeV $\beta$-electron and two
$\gamma$-quanta with the energies $1.173$ and $1.332$~MeV \cite{Knoll}. Metal
shielding and container walls largely stop $\beta$-electrons, while absorbing
only a fraction of $\gamma$-rays. All unabsorbed $\gamma$-rays, even scattered
ones, will eventually leak through the walls and propagate in air. An example
of energy distribution of $\gamma$-rays and electrons outside the container
wall is shown in Fig.\ \ref{Fig.MCNP} reproduced from Ref. \cite{NusiInfr}.%
%TCIMACRO{\FRAME{ftbpFU}{4.6735in}{2.6521in}{0pt}{\Qcb{Results of Monte-Carlo
%simulations (MCNP code) for 1 Curie of $_{27}^{60}$Co showing the energy
%spectrum of electrons (red) and photons (blue) in a 1cm$^{3}$ tally volume
%outside a 0.5 cm thick steel shipping container. Solid lines are the results
%of calculations for the case when electron cutoff energy inside the steel wall
%adjacent to tally volume is set to 1 keV, while dotted lines are for electron
%cutoff energy set to 100 keV. }}{\Qlb{Fig.MCNP}}{Figure}%
%maintain-aspect-ratio TRUE;  display "USEDEF";  valid_file "T";
%width 4.6735in;  height 2.6521in;  depth 0pt;  original-width 4.7338in;
%original-height 2.6743in;  cropleft "0";  croptop "1";  cropright "1";
%cropbottom "0";  tempfilename 'LXYN9F07.wmf';tempfile-properties "XPR";}} }%
%BeginExpansion
%EndExpansion

The energetic $\gamma$-quanta escaped from the container will occasionally
collide with air molecules, experiencing Compton scattering and photoelectric
absorption, as discussed below. Their propagation distance can be estimated
based on the interaction cross-section $\sigma$ given, e.g., in Ref.
\cite{Carron} on Figures 2.50 (d) and (e) for nitrogen and oxygen,
respectively. (One of these very similar figures was reproduced in Ref.
\cite{NusiInfr}.) From these figures one can draw, at least, two conclusions.
First, for energies between 0.1 MeV and 1.2 MeV shown in Fig. 1, the dominant
effect in the production of free electrons is the Compton scattering (this
conclusion agrees with Figures 2.42, 2.43 of Ref. \cite{Carron}). Second, the
interaction cross-section $\sigma$, as the function of energy, gradually
increases from about 0.055 cm$^{2}$/g at 1.2 MeV to about 0.16 cm$^{2}$/g at
0.1 MeV. Correspondingly, the mean free pass of $\gamma$-rays in air
($n_{a}=2n_{m}\approx6\cdot10^{19}\ $atoms/cm$^{3}$) defined as $l=1/(n_{a}%
\sigma)$ varies from $\sim$120\ m for 1.2\ MeV quanta down to about 40\ m for
100\ keV quanta. When the energy of $\gamma$-quanta decreases down to 30 keV
and below, the photoelectric absorption becomes the dominant effect.

It is important that both the Compton scattering and photoelectric absorption
ionize air molecules, producing high-energy electrons within the MeV range.
These primary electrons, which should not be confused with the $\beta
$-electrons directly from radioactivity, cannot propagate too far: they will
be absorbed by air within a few meters \cite{Carron}. During their short
lifetime, however, most of these primary electrons produce thousands of
secondary ones. It is this process that determines the local rate of total
electron production required for the avalanche breakdown \cite{NusinSprangle}.
As discussed above, all unabsorbed $\gamma$-quanta propagate away from the
radioactive source, continuing to produce free electrons. Hence to estimate
the total rate of free-electron production at various distances from the
radioactive source, one needs to analyze the collisional propagation of
$\gamma$-quanta in air. In this paper, we analyze the photon propagation in
the near-source zone, i.e., at distances less or of order of the mean free
pass $l$ estimated above.%

%TCIMACRO{\FRAME{dtbphF}{4.7986in}{2.3523in}{0pt}{}{}{Figure}%
%maintain-aspect-ratio TRUE;  display "USEDEF";  valid_file "T";
%width 4.7986in;  height 2.3523in;  depth 0pt;  original-width 4.8615in;
%original-height 2.3682in;  cropleft "0";  croptop "1";  cropright "1";
%cropbottom "0";  tempfilename 'LXYN9F08.wmf';tempfile-properties "XPR";}} }%
%BeginExpansion
\begin{center}
\includegraphics[
natheight=2.352300in,
natwidth=4.798600in,
height=2.3523in,
width=4.7986in
]%
{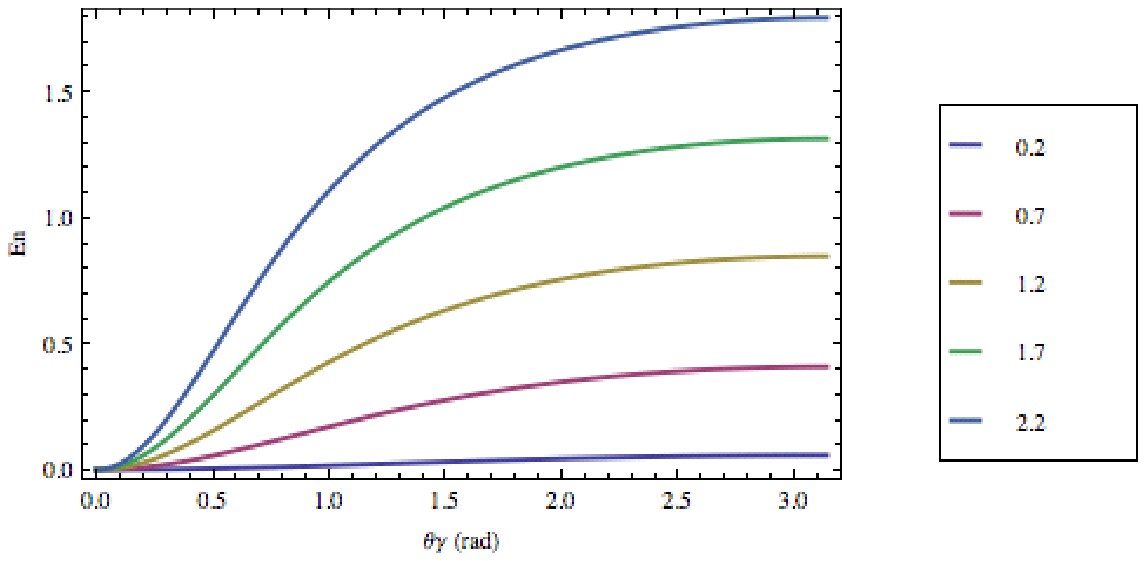}%
\end{center}
%EndExpansion
%TCIMACRO{\FRAME{ftbpF}{4.5467in}{2.2343in}{0pt}{}{}{Figure}%
%maintain-aspect-ratio TRUE;  display "USEDEF";  valid_file "T";
%width 4.5467in;  height 2.2343in;  depth 0pt;  original-width 4.6052in;
%original-height 2.2476in;  cropleft "0";  croptop "1";  cropright "1";
%cropbottom "0";  tempfilename 'LXYN9F09.wmf';tempfile-properties "XPR";}} }%
%BeginExpansion
\begin{figure}
[ptb]
\begin{center}
\includegraphics[
natheight=2.234300in,
natwidth=4.546700in,
height=2.2343in,
width=4.5467in
]%
{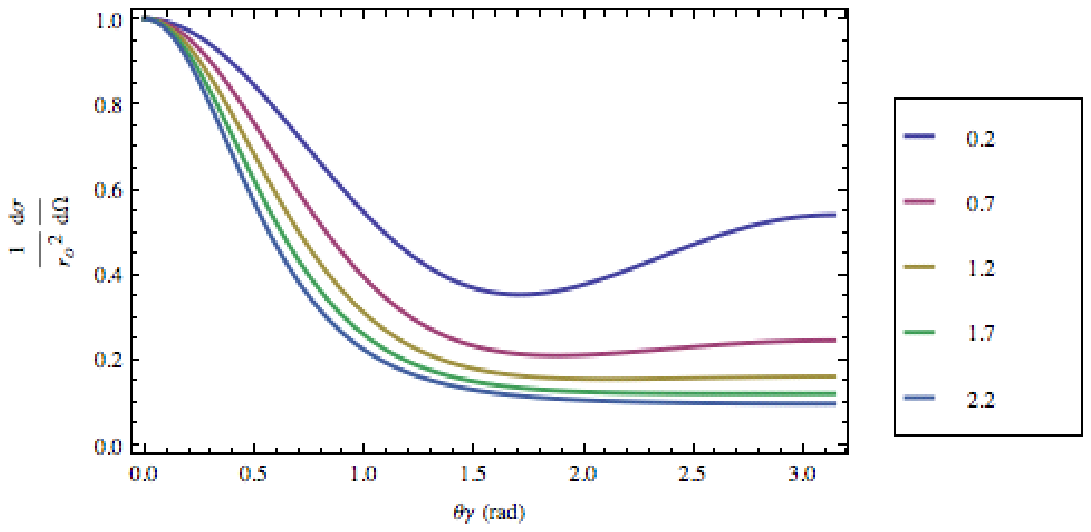}%
\end{center}
\end{figure}
%EndExpansion
%TCIMACRO{\FRAME{ftbpFU}{5.3015in}{2.5962in}{0pt}{\Qcb{Kinetic energy, KN
%differential cross section and angular differential cross-section of gamma
%photons.}}{\Qlb{Fig:KN_figures}}{Figure}%
%maintain-aspect-ratio TRUE;  display "USEDEF";  valid_file "T";
%width 5.3015in;  height 2.5962in;  depth 0pt;  original-width 5.3751in;
%original-height 2.6175in;  cropleft "0";  croptop "1";  cropright "1";
%cropbottom "0";  tempfilename 'LXYN9F0A.wmf';tempfile-properties "XPR";}} }%
%BeginExpansion
\begin{figure}
[ptb]
\begin{center}
\includegraphics[
natheight=2.596200in,
natwidth=5.301500in,
height=2.5962in,
width=5.3015in
]%
{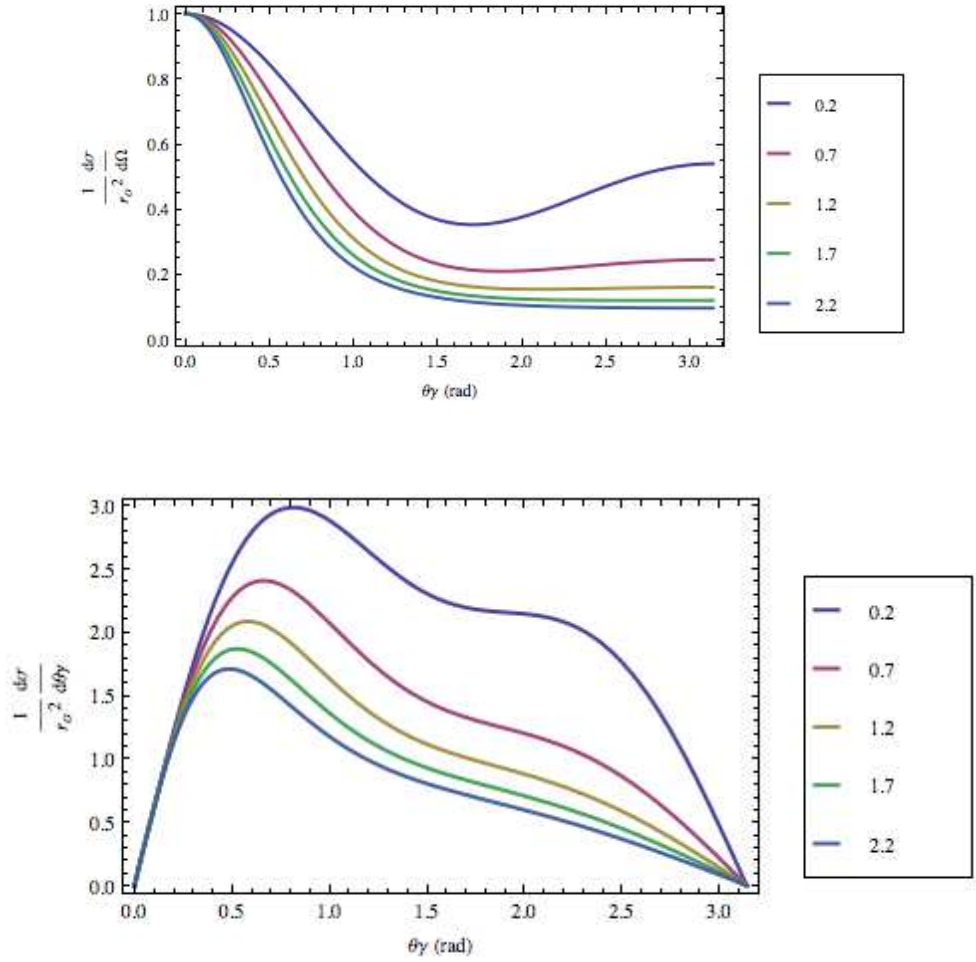}%
\caption{Kinetic energy, KN differential cross section and angular
differential cross-section of gamma photons.}%
\label{Fig:KN_figures}%
\end{center}
\end{figure}
%EndExpansion

\section{Production of free electrons\label{Production of free electrons}}

As propagating $\gamma$-quanta collide with air molecules, producing
high-energy primary electrons, they experience a strong angular scatter and
energy losses. The major electron production takes place when the photon
energy $\mathcal{E}_{\gamma}$ remains between 30\ keV and 1.3\ MeV, i.e., in
the energy range where the Compton scattering dominates. In this range, one
can successfully apply the Klein-Nishina (K-N) theory that describes Compton
scattering of photons by free electrons \cite{Carron}. Indeed, after a typical
collision between the incident photon and an electron bound within an O$_{2}$
or N$_{2}$ molecule, the originally bound electron becomes free and acquires a
significant momentum (energy). If the acquired energy $\mathcal{E}_{\gamma}$
greatly exceeds the electron binding energy then both the scattered photons
and released electrons will have momentum distributions fairly close to those
if the originally bound electron was free and at rest. Furthermore, if the
incident-photon energy is well above the maximum electron binding energy (the
K-shell edge), $\mathcal{E}_{K}\simeq500$\ eV for the air molecules, then the
K-N theory should be applicable to all bound electrons with the exception of
small-angle scattering when the released electrons acquire a small energy,
$\mathcal{E}_{e}\lesssim\mathcal{E}_{K}$ \cite[Section 2.6.2.1]{Carron}.

As described in numerous textbooks, see, e.g., Ref.\ \cite[Section\ 2.6]%
{Carron}, in the frameworks of the K-N theory, the kinetic energy, $T^{\prime
}\equiv c\left(  p_{e}^{2}+m_{e}^{2}c^{2}\right)  ^{1/2}-m_{e}c^{2}%
=\mathcal{E}_{\gamma}-\mathcal{E}_{\gamma}^{\prime}$, of the outgoing electron
and the scalar momentum of the scattered photon, $p_{\gamma}^{\prime}=|\vec
{p}_{\gamma}^{\prime}|=\mathcal{E}_{\gamma}^{\prime}/c$ (for the incident
photon with $p_{\gamma}=|\vec{p}_{\gamma}|=\mathcal{E}_{0}/c$) are given by
\begin{subequations}
\label{T',p'}%
\begin{align}
T^{\prime}  &  \equiv\frac{\alpha^{2}\left(  1-\cos\Theta_{\gamma}\right)
m_{e}c^{2}}{1+\alpha\left(  1-\cos\Theta_{\gamma}\right)  },\label{T'}\\
p_{\gamma}^{\prime}  &  =\left(  \frac{1}{p_{\gamma}}+\frac{1-\cos
\Theta_{\gamma}}{m_{e}c}\right)  ^{-1}, \label{p'}%
\end{align}
while the K-N differential scattering cross-section is given by%
\end{subequations}
\begin{align}
&  \frac{d\sigma_{\mathrm{KN}}}{d\Omega_{\gamma}}=\frac{r_{0}^{2}}{2}\left(
\frac{p_{\gamma}^{\prime}}{p_{\gamma}}\right)  ^{2}\left(  \frac{p_{\gamma}%
}{p_{\gamma}^{\prime}}+\frac{p_{\gamma}^{\prime}}{p_{\gamma}}-\sin^{2}%
\Theta_{\gamma}\right) \nonumber\\
&  =\frac{r_{0}^{2}}{2\left[  1+\alpha\left(  1-\cos\Theta_{\gamma}\right)
\right]  ^{2}}\left[  1+\cos^{2}\Theta_{\gamma}+\frac{\alpha^{2}\left(
1-\cos\Theta_{\gamma}\right)  ^{2}}{1+\alpha\left(  1-\cos\Theta_{\gamma
}\right)  }\right]  . \label{Klein-Nishina}%
\end{align}
In (\ref{T',p'})--(\ref{Klein-Nishina}), $\Theta_{\gamma}$ is the scattering
angle, i.e., the angle between the momenta of the incident and scattered
photons, $\vec{p}_{\gamma}$ and $\vec{p}_{\gamma}^{\prime}$, which is rigidly
related to $p_{\gamma}$ and $p_{\gamma}^{\prime}$,
\begin{equation}
\cos\Theta_{\gamma}=1-\frac{m_{e}c}{p_{\gamma}^{\prime}}+\frac{m_{e}%
c}{p_{\gamma}}, \label{cos_theta_gamma}%
\end{equation}
$\Omega_{\gamma}=(\Theta_{\gamma},\Phi_{\gamma})$ is the solid scattering
angle, $m_{e}$ is the electron rest mass, $c$ is the vacuum speed of light,
$\alpha\equiv\mathcal{E}_{\gamma}/(m_{e}c^{2})=p_{\gamma}/(m_{e}c)$ is the
energy of the incident photon normalized to the electron rest energy and
$r_{0}\approx2.818\times10^{-13}$~$%
%TCIMACRO{\unit{cm}}%
%BeginExpansion
\operatorname{cm}%
%EndExpansion
$ is the classical electron radius.

\begin{figure}
[ptb]
\begin{center}
\includegraphics[
natheight=5.160500in,
natwidth=6.670100in,
height=5.1605in,
width=6.6701in
]%
{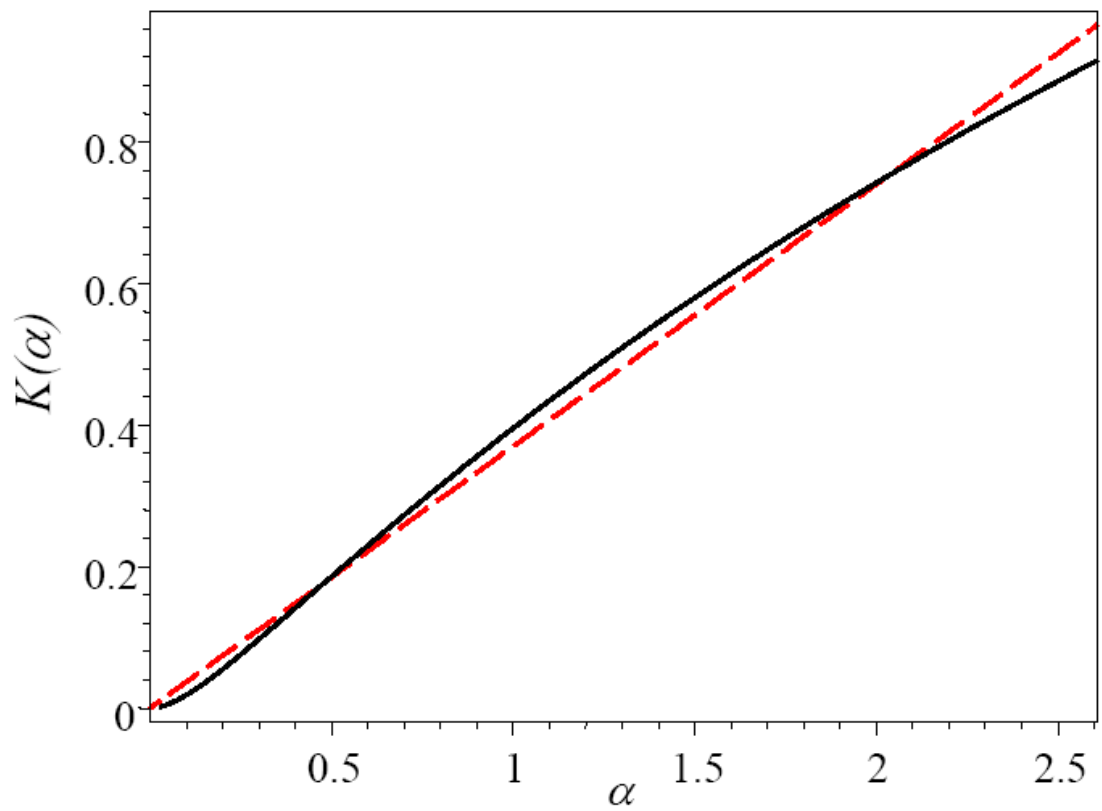}%
\caption{Function $K(\alpha)$ (solid black) and its linear approximation
$0.37\alpha$ (dashed red)}%
\label{Fig:KK}%
\end{center}
\end{figure}

The K-N formula (\ref{Klein-Nishina}) allows one to calculate the total
cross-section and, therefore, the production rate of the primary electrons,
$2Z_{\mathrm{Air}}n_{m}c\sigma_{\mathrm{KN}}$, where $n_{m}=n_{a}%
/2\approx3\cdot10^{19}\ $cm$^{-3}$ is the total density of air molecules and
$Z_{\mathrm{Air}}$ is the mean air atomic number (for the standard air
composition of 79\% N$_{2}$ with $Z_{\text{N}}=7$ and 21\% O$_{2}$ with
$Z_{\text{O}}=8$, we have $Z_{\mathrm{Air}}\approx7.21$). Of prime interest to
us is the production rate, $dN_{e}/dt$, of secondary (knock-on) electrons
which will inevitably end up attaching to neutral molecules, i.e., forming
negative ions \cite{NusinSprangle}. Each primary electron with the kinetic
energy $T$ produces on average $T/(34%
%TCIMACRO{\unit{eV}}%
%BeginExpansion
\operatorname{eV}%
%EndExpansion
)$ secondary electrons \cite{Knoll}--\cite{ValentCurran}. Then, to determine
the total average rate of secondary electron production by photons of a given
initial energy $\alpha$, we should first integrate the K-N differential
cross-section over all solid angles with the weight factor $T^{\prime}$ given
by (\ref{T'}) and (\ref{cos_theta_gamma}),%
\begin{align}
&  \sigma_{T}\equiv\int\frac{d\sigma_{\mathrm{KN}}}{d\Omega_{\gamma}%
}\ T^{\prime}d\Omega_{\gamma}=2\pi\int_{-1}^{1}\frac{d\sigma_{\mathrm{KN}}%
}{d\Omega_{\gamma}}\ T^{\prime}(\Theta_{\gamma})d\left(  \cos\Theta_{\gamma
}\right) \nonumber\\
&  =\pi r_{0}^{2}m_{e}c^{2}\alpha^{2}\int_{-1}^{1}\frac{\left(  1-x\right)
}{\left[  1+\alpha\left(  1-x\right)  \right]  ^{3}}\left[  1+x^{2}%
+\frac{\alpha^{2}\left(  1-x\right)  ^{2}}{1+\alpha\left(  1-x\right)
}\right]  dx, \label{promka}%
\end{align}
and then divide the result by $34%
%TCIMACRO{\unit{eV}}%
%BeginExpansion
\operatorname{eV}%
%EndExpansion
$. The exact calculation of the last integral of (\ref{promka}) yields a
function%
\begin{align*}
&  K(\alpha)\equiv\alpha^{2}\int_{-1}^{1}\frac{\left(  1-x\right)  }{\left[
1+\alpha\left(  1-x\right)  \right]  ^{3}}\left[  1+x^{2}+\frac{\alpha
^{2}\left(  1-x\right)  ^{2}}{1+\alpha\left(  1-x\right)  }\right]  dx\\
&  =\frac{18+102\alpha+186\alpha^{2}+102\alpha^{3}-20\alpha^{4}}%
{3\alpha\left(  1+2\alpha\right)  ^{3}}+\frac{\left(  1+\alpha\right)  \left(
\alpha-3\right)  }{\alpha^{2}}\ln\left(  1+2\alpha\right)  ,
\end{align*}
shown in Fig.\ \ref{Fig:KK} by the black solid line. Note that the function
$K(\alpha)$ is proportional to the average \textquotedblleft absorption cross
section\textquotedblright\ \cite[Eq. (8e-28)]{AmHandbook},%
\begin{equation}
\sigma_{a}=\frac{\pi r_{0}^{2}K(\alpha)}{\alpha}. \label{relation}%
\end{equation}
%TCIMACRO{\FRAME{ftbpFU}{6.6701in}{5.1605in}{0pt}{\Qcb{Function $K(\alpha)$
%(solid black) and its linear approximation $0.37\alpha$ (dashed red)}%
%}{\Qlb{Fig:KK}}{Figure}{\special{ language "Scientific Word";
%type "GRAPHIC";  maintain-aspect-ratio TRUE;  display "USEDEF";
%valid_file "T";  width 6.6701in;  height 5.1605in;  depth 0pt;
%original-width 8.4618in;  original-height 6.5388in;  cropleft "0";
%tempfilename 'LXYN9F0B.wmf';tempfile-properties "XPR";}} }%
%BeginExpansion
%EndExpansion
In the energy range of interest, $\mathcal{E}_{\gamma}\leq1.332\
%TCIMACRO{\unit{MeV}}%
%BeginExpansion
\operatorname{MeV}%
%EndExpansion
$, i.e., $\alpha\leq\mathcal{\alpha}_{\max}\approx2.61$, the complex function
$K(\alpha)$ can be quite accurately approximated by the dependence
$0.37\alpha$ ($\sigma_{a}=0.37\pi r_{0}^{2}\approx9.23\times10^{-26}%
%TCIMACRO{\unit{cm}}%
%BeginExpansion
\operatorname{cm}%
%EndExpansion
^{2}$) shown in Fig.\ \ref{Fig:KK} by the red dashed line. This approximation
allows us to describe the total average rate of secondary electron production
by a simple formula:%
\begin{align}
\frac{dN_{e}}{dt}  &  =2Z_{\mathrm{Air}}n_{m}c\sigma_{T}n_{\gamma}%
\simeq2.32Z_{\mathrm{Air}}r_{0}^{2}n_{m}c\left(  \frac{m_{e}c^{2}}{34%
%TCIMACRO{\unit{eV}}%
%BeginExpansion
\operatorname{eV}%
%EndExpansion
}\right)  \int_{0}^{\infty}\alpha F_{\gamma}(\alpha)d\alpha\nonumber\\
&  \approx2.09\times10^{10}%
%TCIMACRO{\unit{s}}%
%BeginExpansion
\operatorname{s}%
%EndExpansion
^{-1}\int_{0}^{\mathcal{\infty}}\alpha F_{\gamma}(\alpha)d\alpha,
\label{dN_e/dt}%
\end{align}
where $F_{\gamma}(\alpha)$ is the photon energy distribution normalized to the
total photon density, $n_{\gamma}=\int F_{\gamma}(\alpha)d\alpha$, and
expressed in terms of the full momentum distribution function $f_{\gamma}%
(\vec{p}_{\gamma})$ (Sect.\ \ref{Photon_Kinetics}) as $F_{\gamma}%
(\alpha)=m_{e}^{3}c^{3}\alpha^{2}\int f_{\gamma}(p_{\gamma},\Omega_{\gamma
})d\Omega_{\gamma}$. Thus, to evaluate the average rate of secondary electron
production as a function of the distance from the radiation source, one needs
to determine the photon energy distribution along with its distance
variations. This requres a collisional kinetic theory for propagating photons.

\section{Photon kinetics in spherically symmetric
approximation\label{Photon_Kinetics}}

Radiated and scattered $\gamma$-quanta can be treated as a gas of incoherent
and unpolarized photons. For photon energies above 30 keV, this gas is prone
mainly to Compton scattering, while at lower energies the photoelectric
absorption starts dominating. Both processes result in air molecule
ionization, i.e., production of free electrons. The coherent Rayleigh
scattering is at least an order of magnitude smaller and can be neglected.
Also, the process inverse to photon absorption, viz., photon production during
collisions of free electrons with air molecules, in the range of electron
energies we are dealing with ($\leq1.3$ MeV), is negligibly small. Therefore,
we will assume the spatial and momentum distribution of photons to be
independent of any feedback from the free electrons.

We will treat individual $\gamma$-quanta kinetically as classical particles
localized in both real and momentum spaces. The photon momentum distribution,
$f_{\gamma}(\vec{p}_{\gamma})$, may have a temporal ($t$) and spatial
($\vec{r}$) dependences. Since photons propagate and scatter practically
instantaneously we will ingore any temporal dependence. Being concerned mainly
with distances from the radioactive material much longer than the container
size $a$, we will treat this material as a point-like source of $\gamma
$-quanta. Ignoring possible effects of the ground, air inhomogeneity and
container anisotropy, we will assume spherically symmetric propagation of photons.

\begin{figure}
[ptb]
\begin{center}
\includegraphics[
natheight=5.179100in,
natwidth=7.121600in,
height=5.1791in,
width=7.1216in
]%
{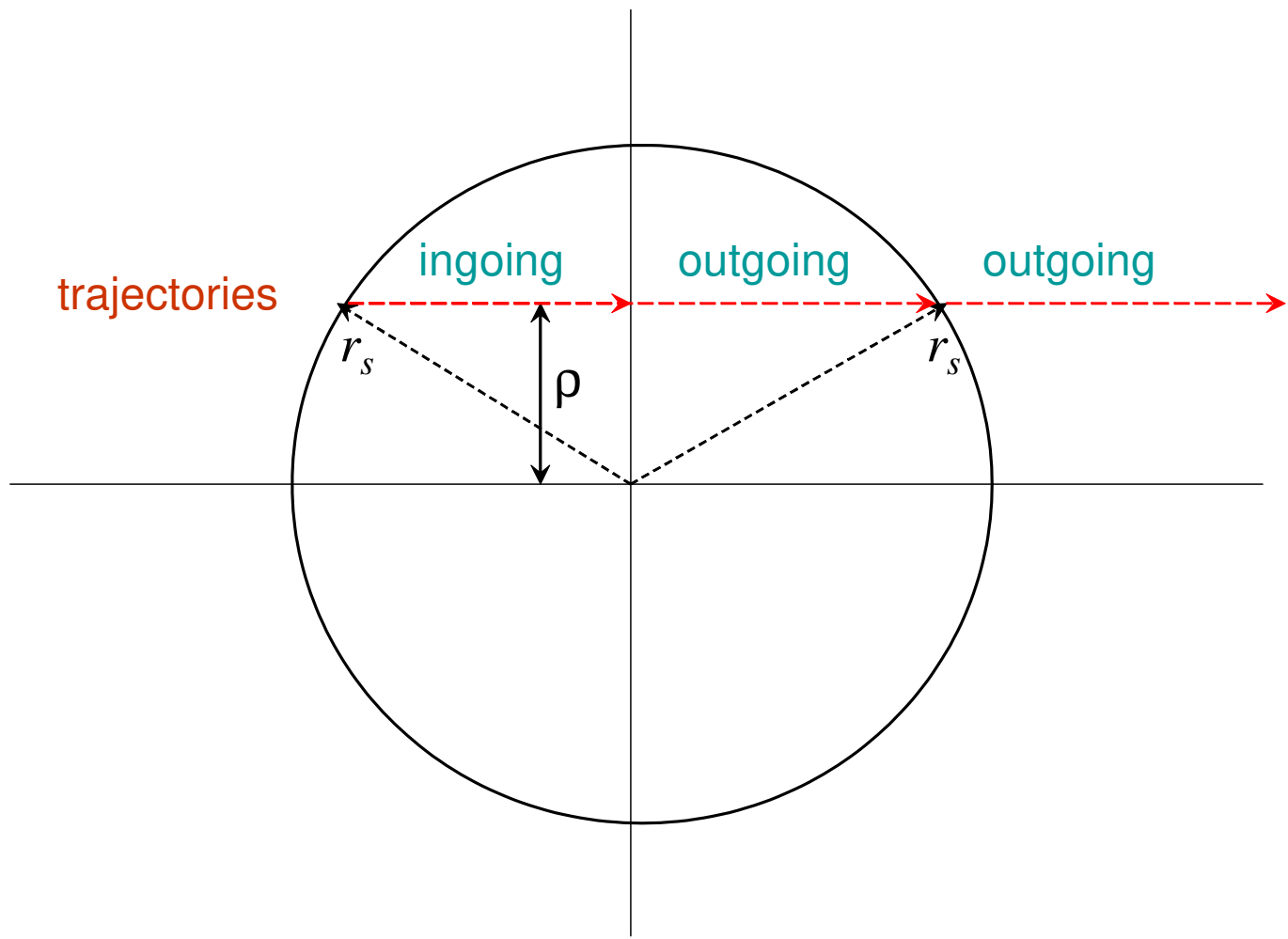}%
\caption{Schematics of photon trajectories.}%
\label{Fig:Trajectories}%
\end{center}
\end{figure}

In the spherically symmetric geometry, the photon momentum distribution
depends on one spatial variable, viz., the distance $r$ of a localized photon
from the point-like source, and two momentum components: the absolute value of
the momentum, $p_{\gamma}$, and the momentum polar angle with respect to the
radius vector at the given location, $\theta_{\gamma}$ (the latter should not
be confused with the angle between the scattered and incident photons,
$\Theta_{\gamma}$). Thus the photon momentum distribution is a function of
three variables, $f_{\gamma}\left(  r,p_{\gamma},\theta_{\gamma}\right)  $.
Since the angle $\theta_{\gamma}$ is not invariant with respect to the
collisionless motion, it is more convenient to pass from $\theta_{\gamma}$ to
the \textquotedblleft impact parameter\textquotedblright\ $\rho$,
\begin{equation}
\rho\equiv r\sin\theta_{\gamma},\qquad\cos\theta_{\gamma}(\rho)=\pm\left(
1-\frac{\rho^{2}}{r^{2}}\right)  ^{1/2}, \label{rho}%
\end{equation}
equal to the shortest distance from the radioactive source ($r=0$) to the
straight-line trajectory of a freely propagating photon. The parameter $\rho$
is proportional to the photon angular momentum and remains invariant during
the free photon propagation. It can be changed only during brief collisions
that lead to Compton scattering. For any given positive $\rho$, we distinguish
between two groups of photons with different signs of $\cos\theta_{\gamma
}(\rho)$ in (\ref{rho}): the \textquotedblleft outgoing\textquotedblright%
\ photons with the `$+$' sign (corresponding to $dr/dt>0$) and
\textquotedblleft ingoing\textquotedblright\ photons with the `$-$' sign
($dr/dt<0$), as illustrated by Figure\ \ref{Fig:Trajectories}. We will denote
the two corresponding distribution functions by $f_{\gamma}^{\pm}$.%

%TCIMACRO{\FRAME{ftbpFU}{7.1216in}{5.1791in}{0pt}{\Qcb{Schematics of photon
%trajectories.}}{\Qlb{Fig:Trajectories}}{Figure}%
%maintain-aspect-ratio TRUE;  display "USEDEF";  valid_file "T";
%width 7.1216in;  height 5.1791in;  depth 0pt;  original-width 9.6867in;
%original-height 7.0338in;  cropleft "0";  croptop "1";  cropright "1";
%cropbottom "0";  tempfilename 'LOYE9B01.wmf';tempfile-properties "XNPR";}} }%
%BeginExpansion
%EndExpansion

In terms of $f_{\gamma}^{\pm}\left(  r,p_{\gamma},\rho\right)  $, the total
photon number density, $n_{\gamma}$, and radial flux density, $g_{\gamma
}\left(  r\right)  $, are given by
\begin{subequations}
\label{ng_ph}%
\begin{align}
n_{\gamma}\left(  r\right)   &  \equiv\sum_{\pm}\int f_{\gamma}^{\pm}%
d^{3}p_{\gamma}=\frac{2\pi}{r}\sum_{\pm}\int_{0}^{\infty}p_{\gamma}%
^{2}dp_{\gamma}\int_{0}^{r}\frac{f_{\gamma}^{\pm}\left(  r,p_{\gamma}%
,\rho\right)  \rho d\rho}{\left(  r^{2}-\rho^{2}\right)  ^{1/2}}%
,\label{ng_ph_n}\\
g_{\gamma}\left(  r\right)   &  \equiv c\sum_{\pm}\int\left(  \cos
\theta_{\gamma}\right)  f_{\gamma}^{\pm}d^{3}p_{\gamma}=\frac{2\pi c}{r^{2}%
}\sum_{\pm}\left(  \pm\int_{0}^{\infty}p_{\gamma}^{2}dp_{\gamma}\int_{0}%
^{r}f_{\gamma}^{\pm}\left(  r,p_{\gamma},\rho\right)  \rho d\rho\right)  .
\label{ng_ph_g}%
\end{align}

The momentum distribution function of freely propagating $\gamma$-quanta with
occasional spiky collisions satisfies the Boltzmann kinetic equation derived
in Appendix \ref{Deriving kinetic equation},%

\end{subequations}
\begin{equation}
\pm c\left(  1-\frac{\rho^{2}}{r^{2}}\right)  ^{1/2}\frac{\partial f_{\gamma
}^{\pm}}{\partial r}=\hat{L}_{r}f_{\gamma}^{\prime\pm}-\nu(p_{\gamma
})f_{\gamma}^{\pm}, \label{kinEq}%
\end{equation}
where the left-hand side describes the collisionless propagation of photons,
while the right-hand side (RHS) describes their collisions with air molecules.

The first term in the RHS\ of (\ref{kinEq}) describes the kinetic arrival\ of
photons to the elementary phase volume $d^{3}p_{\gamma}$ around given $\vec
{p}_{\gamma}$ due to collisional scattering of incident photons from
$d^{3}p_{\gamma}^{\prime}$ around $\vec{p}_{\gamma}^{\prime}$,%
\begin{align}
&  \hat{L}_{r}f_{\gamma}^{\prime\pm}=\frac{2Z_{\mathrm{air}}n_{m}r_{0}%
^{2}m_{e}c^{2}}{p_{\gamma}^{2}}\int_{p_{\gamma}}^{\infty}dp_{\gamma}^{\prime
}\left(  \frac{p_{\gamma}}{p_{\gamma}^{\prime}}+\frac{p_{\gamma}^{\prime}%
}{p_{\gamma}}-\sin^{2}\Theta_{\gamma}\right) \nonumber\\
&  \times\int_{C^{-}}^{C^{+}}\frac{f_{\gamma}^{\prime}\left(  p_{\gamma
}^{\prime},\cos\theta_{\gamma}^{\prime}\right)  d\left(  \cos\theta_{\gamma
}^{\prime}\right)  }{\sqrt{S\left(  \cos\Theta_{\gamma},\cos\theta_{\gamma
},\cos\theta_{\gamma}^{\prime}\right)  }}, \label{I_arr}%
\end{align}
where%
\begin{align}
&  S\left(  \cos\Theta_{\gamma},\cos\theta_{\gamma},\cos\theta_{\gamma
}^{\prime}\right)  =\left(  C^{+}-\cos\theta_{\gamma}^{\prime}\right)  \left(
\cos\theta_{\gamma}^{\prime}-C^{-}\right)  ,\label{S}\\
&  C^{\pm}\equiv\cos\left(  \Theta_{\gamma}\mp\theta_{\gamma}\right)
=\cos\Theta_{\gamma}\cos\theta_{\gamma}\pm\left(  1-\cos^{2}\Theta_{\gamma
}\right)  ^{1/2}\sin\theta_{\gamma}\nonumber
\end{align}
and $\Theta_{\gamma}$ is the angle of Compton scattering from $\vec{p}%
_{\gamma}^{\prime}$ to $\vec{p}_{\gamma}$,%
\begin{equation}
\cos\Theta_{\gamma}(p_{\gamma},p_{\gamma}^{\prime})=1-\frac{m_{e}c}{p_{\gamma
}}+\frac{m_{e}c}{p_{\gamma}^{\prime}} \label{cos_Theta}%
\end{equation}
(here $p_{\gamma}\leq p_{\gamma}^{\prime}$). In (\ref{I_arr}), we keep
$\theta_{\gamma}$, $\theta_{\gamma}^{\prime}$ for easier integration over the
angular variables, but before appplying $\partial/\partial r$ in the LHS of
(\ref{kinEq}), $\cos\theta_{\gamma}$ should be expressed in terms of $\rho$
using (\ref{rho}).

The last term in the RHS of (\ref{kinEq})\ describes the total kinetic
departure\ of $\vec{p}_{\gamma}$-photons due to their scattering with the
total collision frequency, $\nu(p_{\gamma}=m_{e}c\alpha)$, proportional to the
total K-N cross-section \cite{Carron},
\begin{align}
&  \nu=2cn_{m}Z_{\mathrm{air}}\sigma_{\mathrm{KN}}\nonumber\\
&  =4\pi n_{m}r_{0}^{2}cZ_{\mathrm{air}}\left[  \frac{2+8\alpha+9\alpha
^{2}+\alpha^{3}}{\alpha^{2}\left(  1+2\alpha\right)  ^{2}}-\frac{\left(
2+2\alpha-\alpha^{2}\right)  \ln\left(  1+2\alpha\right)  }{2\alpha^{3}%
}\right]  . \label{nu_KN(p)}%
\end{align}

Equation (\ref{kinEq}) should be supplemented by the boundary conditions. In
accord with (\ref{ng_ph_g}), at a small distance $r$ that satisfies $a\ll r\ll
l=c/\nu$, the photon distribution function radiated by the point-like
radioactive source can be written as
\begin{align}
\left.  f_{\gamma}^{+}\left(  p_{\gamma},\rho\right)  \right\vert
_{r\rightarrow0}  &  =\sum_{s=1}^{k}\left.  f_{\gamma s}^{+}\left(  p_{\gamma
},\rho\right)  \right\vert _{r\rightarrow0},\nonumber\\
\left.  f_{\gamma s}^{+}\left(  p_{\gamma},\rho\right)  \right\vert
_{r\rightarrow0}  &  =\frac{I_{\gamma}^{(0)}\delta\left(  \cos\theta_{\gamma
}-1\right)  \delta\left(  p_{\gamma}-p_{s}\right)  }{8\pi^{2}cr^{2}p_{s}^{2}%
}=\frac{I_{\gamma}^{(0)}\delta\left(  \rho\right)  \delta\left(  p_{\gamma
}-p_{s}\right)  }{8\pi^{2}c\rho p_{s}^{2}}, \label{source_beam}%
\end{align}
where $I_{\gamma}^{(0)}=3.7\times10^{10}%
%TCIMACRO{\unit{s}}%
%BeginExpansion
\operatorname{s}%
%EndExpansion
^{-1}Q\left(  \mathrm{%
%TCIMACRO{\unit{Ci}}%
%BeginExpansion
\operatorname{Ci}%
%EndExpansion
}\right)  $ is the total number of radioactive decays per unit time,
$\sum_{s=1}^{k}$ denotes summation over all kinds of primary photons radiated
during one disintegration of the radioactive material (usually, $k=1$,$2$,$3$)
and $\delta\left(  p_{\gamma}-p_{s}\right)  $ expresses the fact that all
radiated photons have discrete energies $\mathcal{E}_{s}=cp_{s}$ (e.g., for
$_{27}^{60}$Co: $k=2$, $\mathcal{E}_{1}=1.332$\ $%
%TCIMACRO{\unit{MeV}}%
%BeginExpansion
\operatorname{MeV}%
%EndExpansion
$ and $\mathcal{E}_{2}=1.173$\ $%
%TCIMACRO{\unit{MeV}}%
%BeginExpansion
\operatorname{MeV}%
%EndExpansion
$). Equation (\ref{source_beam}) represents the inner boundary condition at
$r\rightarrow0$ applicable to the outgoing photons. For the ingoing photons,
we use the natural condition of $f_{\gamma}^{-}\left(  p_{\gamma},\rho\right)
\rightarrow0$ at $r\rightarrow\infty$. Between consecutive collisions, some
ingoing photons pass the minimum distance $r=\rho$ and become outgoing. This
yields the matching condition between the two groups:%
\begin{equation}
f_{\gamma}^{+}\left(  r,p_{\gamma},\rho=r\right)  =f_{\gamma}^{-}\left(
r,p_{\gamma},\rho=r\right)  . \label{matching}%
\end{equation}
Further, all collisional processes lead only to decreasing photon energies, so
there are no photons with energies exceeding the initial ones, $f_{\gamma}%
^{+}\left(  p>\max(p_{s})\right)  =0$. Lastly, we assume that at $\rho=0$
($\cos\theta_{\gamma}=\pm1$) all momentum distributions behave regularly.

The RHS of (\ref{kinEq}) satisfies automtically the condition $\int
_{0}^{\infty}p_{\gamma}^{2}dp_{\gamma}\int_{0}^{r}(\hat{L}_{r}f_{\gamma
}^{\prime\pm}-\nu f_{\gamma}^{\prime\pm})\left(  r^{2}-\rho^{2}\right)
^{-1/2}\rho d\rho=0$, which provides in accord with (\ref{ng_ph_g})
conservation of the total photon flux through any surface surrounding the
radiation source. This flux conservation is due to the fact that kinetic
equation (\ref{kinEq}) includes only photon scattering. Photoelectric
absorption and other possible photon losses can also be included in
$\nu(p_{\gamma})$ by a simple addition to the RHS\ of (\ref{nu_KN(p)}). Those
effects would break the total flux conservation.

Kinetic equation (\ref{kinEq}) is solved analytically in
Appendix\ \ref{Solving kinetic equation}. The complete solution is given by an
infinite series of multiply scattered photons (\ref{f=}), where $f_{0}%
^{+}=\left(  f_{\gamma}^{+}\right)  _{\mathrm{prim}}$ is given by
(\ref{F_prim}), $f_{n\geq1}^{\pm}$ is determined by recurrence relations
(\ref{recur_relation}) with the linear integral operator $\hat{L}_{r}$ defined
by (\ref{I_arr}). This general solution, however, involves an increasing
number of nested integrations which for $n\geq2$ are difficult to calculate.
Below we restrict our treatment to distances $r$ from the radiation source
located within the mean free path for 1.2\ MeV $\gamma$-quanta, $r\lesssim
l_{s}\equiv c/\nu(p_{s})\sim120\
%TCIMACRO{\unit{m}}%
%BeginExpansion
\operatorname{m}%
%EndExpansion
$ (the \textquotedblleft near zone\textquotedblright). At these distances, one
might expect two dominant photon groups: the primary $\gamma$-quanta beams
($n=0$) and singly scattered photons ($n=1$). The momentum distribution for
singly scattered photons is obtained in Appendix\ \ref{solution for n=1}.
However, a mere cutoff of the infinite series (\ref{f=}) to its first two
terms, $f_{0}^{+}$ and $f_{1}^{\pm}$ given by (\ref{external_p>p_gamma}%
)--(\ref{A_int+-}), becomes less satisfactory at distances $r\sim l_{s}$
because further scattering of singly scattered photons breaks the total flux
conservation. Below we modify the two-group description of photons to make it
exactly flux conserving.

\section{Photon energy distribution in the near zone}

To calculate the total electron production rate $dN_{e}/dt$ given by
(\ref{dN_e/dt}), we need to determine the spatially dependent photon energy
distribution at various distances from the radioactive source, $F_{\gamma
}(\alpha)$, where $\alpha=p_{\gamma}/(m_{e}c)$ is the normalized energy. In
this section, however, it is more convenient to use%
\begin{equation}
F(p_{\gamma})=p_{\gamma}^{2}\int f_{\gamma}(p_{\gamma},\Omega_{\gamma}%
)d\Omega_{\gamma}=\frac{2\pi p_{\gamma}^{2}}{r}\sum_{\pm}\int_{0}^{r}%
\frac{f_{\gamma}^{\pm}\left(  r,p_{\gamma},\rho\right)  \rho d\rho}{\left(
r^{2}-\rho^{2}\right)  ^{1/2}}, \label{F(alpha)}%
\end{equation}
which differs from by the normalization factor following from $dn_{\gamma
}=F(p_{\gamma})dp_{\gamma}=F(\alpha)d\alpha$: $F_{\gamma}(\alpha
)=m_{e}cF(p_{\gamma})$. In terms of $F(p_{\gamma})$, (\ref{dN_e/dt}) becomes%
\begin{equation}
\left(  \frac{dN_{e}}{dt}\right)  _{\mathrm{prim}}\approx2.09\times10^{10}%
%TCIMACRO{\unit{s}}%
%BeginExpansion
\operatorname{s}%
%EndExpansion
^{-1}\frac{1}{m_{e}c}\int_{0}^{\infty}p_{\gamma}F(p_{\gamma})dp_{\gamma}.
\label{dN_e/dt_via_p}%
\end{equation}

\subsection{Primary photons}

For the primary photons, we have the solution (\ref{F_prim}) valid for
arbitrary distances from the radiation source. The exponential factors
$e^{-r/l_{s}}$ there describe the beam attenuation due to Compton scattering.
For the corresponding $F(p_{\gamma})$, photon density $\left(  n_{\gamma
}\right)  _{\mathrm{prim}}$, total flux $\left(  I_{\gamma}\right)
_{\mathrm{prim}}$ and the electron production rate $\left(  dN_{e}/dt\right)
_{\mathrm{prim}}$, we obtain simple formulas%
\begin{align}
\left[  F(p_{\gamma})\right]  _{\mathrm{prim}}  &  =\frac{I_{\gamma}^{(0)}%
}{4\pi cr^{2}}\sum_{s=1}^{k}\delta\left(  p_{\gamma}-p_{s}\right)
e^{-r/l_{s}},\qquad\left(  n_{\gamma}\right)  _{\mathrm{prim}}=\frac{\left(
I_{\gamma}\right)  _{\mathrm{prim}}}{4\pi cr^{2}}=\frac{I_{\gamma}^{(0)}}{4\pi
cr^{2}}\sum_{s=1}^{k}e^{-r/l_{s}},\nonumber\\
\left(  \frac{dN_{e}}{dt}\right)  _{\mathrm{prim}}  &  \approx2.09\times
10^{10}%
%TCIMACRO{\unit{s}}%
%BeginExpansion
\operatorname{s}%
%EndExpansion
^{-1}\frac{I_{\gamma}^{(0)}}{4\pi cr^{2}}\sum_{s=1}^{k}\frac{p_{s}}{m_{e}%
c}\ e^{-r/l_{s}}\nonumber\\
&  \approx4.017\times10^{5}%
%TCIMACRO{\unit{s}}%
%BeginExpansion
\operatorname{s}%
%EndExpansion
^{-1}%
%TCIMACRO{\unit{cm}}%
%BeginExpansion
\operatorname{cm}%
%EndExpansion
^{-3}\frac{I_{\gamma}^{(0)}(%
%TCIMACRO{\unit{Ci}}%
%BeginExpansion
\operatorname{Ci}%
%EndExpansion
)}{\left[  r(%
%TCIMACRO{\unit{m}}%
%BeginExpansion
\operatorname{m}%
%EndExpansion
)\right]  ^{2}}\sum_{s=1}^{k}\mathcal{E}_{s}(%
%TCIMACRO{\unit{MeV}}%
%BeginExpansion
\operatorname{MeV}%
%EndExpansion
)e^{-r/l_{s}}. \label{dN_e/dt_prim}%
\end{align}

\subsection{Scattered photons: extended $n=1$ solution}

Each primary beam of radiated photons denoted by $s$ produces its own set of
scattered photons characterized by the corresponding partial distribution
function $f_{\gamma s}^{+}$. Bearing in mind that the distribution of all
photons is given by $f_{\gamma}^{+}=\sum_{s=1}^{k}f_{\gamma s}^{+}$ and
\begin{equation}
F(p_{\gamma})=\sum_{s=1}^{k}F_{s}(p_{\gamma}),\qquad F_{s}(p_{\gamma}%
)=\frac{2\pi p_{\gamma}^{2}}{r}\sum_{\pm}\int_{0}^{r}\frac{f_{\gamma s}^{\pm
}\left(  r,p_{\gamma},\rho\right)  \rho d\rho}{\left(  r^{2}-\rho^{2}\right)
^{1/2}} \label{summa_F}%
\end{equation}
below we consider separately the partial distributions $f_{\gamma s}^{+}$ and
the corresponding partial moments.

As a primary beam of photons leaves the radiation source it starts producing
scattered photons with the rate, which in the close proximity to the radiation
source, $r\ll l_{s}$, grows linearly with distance. Convolved with the
$1/r^{2}$ spherical divergence, the population of scattered photons becomes
proportional to $1/r$. Our simulations show that this spatial dependence for
the energy distribution of scattered photons roughly holds through the entire
near zone, $r\lesssim l_{s}$. Until the low-energy photoelectric absorption
becomes important, a partial photon flux of all related photons remains
constant, $I_{\gamma s}=I_{\gamma}^{(0)}$. Each partial primary-photon beam
$s$ decays exponentially due to Compton scattering, $\left(  I_{\gamma
s}\right)  _{\mathrm{prim}}=I_{\gamma}^{(0)}e^{-r/l_{s}}$, so that at
$r\geq0.7l_{s}$ the corresponding flux of scattered photons, $I_{\gamma}%
^{(0)}\left(  1-e^{-r/l_{s}}\right)  $, dominates. The corresponding photon
density, $\left(  n_{\gamma}\right)  _{s}=I_{\gamma}^{(0)}\left(
1-e^{-r/l_{s}}\right)  /\langle v_{rs}\rangle$, where $\langle v_{rs}\rangle$
is the mean radial velocity of scattered photons, starts dominating noticeably
closer to the radiation source due to the more isotropic momentum distribution
($\langle v_{rs}\rangle_{\mathrm{scat}}<c$) compared to that of the almost
perfectly directional primary beam ($\langle v_{rs}\rangle_{\mathrm{prim}%
}\approx c$). The same pertains to the electron production rate, albeit the
distance where $dN_{e}/dt$ produced by scattered photons starts dominating at
a larger distance $r$ than that for the photon density due to the weighting
factor $p_{\gamma}$ in the integrand of (\ref{dN_e/dt_via_p}). (This factor is
biased to higher energies where scattered photons momentum distribution is
more directional.) Thus electron production due to scattered photons may
become important deeply in the near zone, $r\lesssim l_{s}$.

We simplify the treatment of scattered photons in the near zone as follows. We
will consider all scattered photons, $n\geq1$, as one unified group and extend
for this group the singly-scattered ($n=1$) photon momentum distribution given
by (\ref{external_p>p_gamma}) and (\ref{internal_p<p_gamma}) with one simple
modification. In the exponents of the attenuation factors (\ref{A_ext+}) and
(\ref{A_int+-}), we will retain the terms $r_{s}/l_{s}$ responsible for the
primary beam attenuation, but will ignore $[\pm(r_{s}^{2}-\rho^{2})^{1/2}%
\pm(r^{2}-\rho^{2})^{1/2}]/l\left(  p_{\gamma}\right)  $. The reason is that
the latter terms describe the loss of $n=1$ photons due to their further
scattering. For the entire unified group of all scattered photons, this
represents just a particle redistribution and hence must be ignored. This will
provide automatically the total photon flux conservation.

The extension of the $n=1$ solution to all scattered photons will introduce
two inaccuracies. On the one hand, multiply scattered photons decrease their
energy as $n$ increases, so that photons with $n\geq2$ leak into the
lower-energy range $\mathcal{E<E}_{\min}^{(1)}=\mathcal{E}_{s}[1+2p_{s}%
/(m_{e}c)]\simeq0.2\
%TCIMACRO{\unit{MeV}}%
%BeginExpansion
\operatorname{MeV}%
%EndExpansion
$, as discussed in the end of Appendix
section\ \ref{General solution of multi-scattered photons}. This reduces the
photon population in the regular $n=1$ energy range, $\mathcal{E}_{\min}%
^{(1)}\leq\mathcal{E}\leq\mathcal{E}_{s}$. Due to the weighting factor
$p_{\gamma}$ in the integrand of (\ref{dN_e/dt_via_p}), this inaccuracy in the
extended $n=1$ solution would lead to overestimated $dN_{e}/dt$. On the other
hand, multiply scattered photons have a more isotropic angular distribution in
the momentum space that leads to the reduced average photon radial velocity
and, hence, to the increased photon density. As a result, the more directional
extended $n=1$ solution would, on the contrary, underestimate $dN_{e}/dt$. We
expect the two inaccuracies to partially balance each other. As discussed
below by comparing our analytical results with computer simulations, our
simplified analytical model describes well the spatial dependence of the
photon energy distribution within $\mathcal{E}_{\min}^{(1)}\leq\mathcal{E}%
\leq\mathcal{E}_{s}$, practically in the entire near zone ($r<100\
%TCIMACRO{\unit{m}}%
%BeginExpansion
\operatorname{m}%
%EndExpansion
$). Thus it can be employed for reliable estimates of the electron production
rate in the near zone.

The extended $n=1$ solution is even simpler than the original one because
(\ref{A_ext+}) and (\ref{A_int+-}) reduce to a common attenuation factor,
$A_{\mathrm{ext}}^{\pm}=A_{\mathrm{int}}^{\pm}=\exp\left(  -r_{s}%
/l_{s}\right)  $. One can easily check that the modified solution conserves
exactly the total photon flux through any closed surface surrounding the
radiation source. Indeed, the contributions of the ingoing and outgoing
photons at $r<r_{s}$ (see Fig.\ \ref{Fig:Trajectories}) cancel each other. The
remaining partial fluxes of the outgoing photons in the outer sphere yield
$I_{\gamma}^{(0)}\left(  1-e^{-r/l_{s}}\right)  $, in full agreement with the
flux conservation. Unfortunately, the contributions from the ingoing and
outgoing photons within $r<r_{s}$ to the photon energy distribution and
$dN_{e}/dt$ have no opposite signs and add rather than cancel. This leads to
more complicated calculations outlined below.

Using the modified $n=1$ solution, we calculate now the energy distribution of
scattered photons. According to the above, the partial momentum distribution
assoiciated with a beam $s$ is
\begin{equation}
f_{\gamma s}^{+}=\frac{Z_{\mathrm{air}}I_{\gamma}^{(0)}n_{m}r_{0}^{2}%
m_{e}r_{s}}{4\pi p_{\gamma}^{2}\rho^{2}p_{s}^{2}}\exp\left(  -\ \frac
{r_{s}(\rho,p_{\gamma},p_{s})}{l_{s}}\right)  , \label{f_simpl}%
\end{equation}
where
\begin{equation}
l_{s}\equiv l(p_{s})=\frac{c}{\nu(p_{s})}, \label{l_s}%
\end{equation}
and $r_{s}(\rho,p_{\gamma},p_{s})=\rho/L^{1/2}$, introduced in (\ref{r_s}),
describes the distance where the scattering of primary photons took place.
Introducing $L(p_{\gamma},p_{s})=\sin^{2}\Theta_{\gamma}$, as in
(\ref{sin^2Theta}), we obtain%
\begin{equation}
F_{s}(p_{\gamma})=\frac{Z_{\mathrm{air}}I_{\gamma}^{(0)}n_{m}r_{0}^{2}m_{e}%
}{2rp_{s}^{2}}\left(  \frac{p_{\gamma}}{p_{s}}+\frac{p_{s}}{p_{\gamma}%
}-L\right)  \left\{
\begin{array}
[c]{ccc}%
S_{\mathrm{for}} & \text{if} & p_{\gamma}\geq p_{\mathrm{cr}s}=\frac{p_{s}%
}{1+p_{s}/(m_{e}c)}\\
S_{\mathrm{back}} & \text{if} & p_{\gamma}<p_{\mathrm{cr}s}=\frac{p_{s}%
}{1+p_{s}/(m_{e}c)}%
\end{array}
\right.  , \label{F_prome}%
\end{equation}
where the boundary $p_{\gamma}=p_{\mathrm{cr}s}$ corresponds to $L=1$
($\Theta_{\gamma}=90^{\circ}$) and
\begin{subequations}
\label{S_}%
\begin{align}
S_{\mathrm{for}}  &  \equiv\frac{1}{\sqrt{L}}\int_{0}^{rL^{1/2}}\frac
{\exp(-r_{s}/l_{s})d\rho}{\left(  r^{2}-\rho^{2}\right)  ^{1/2}}%
,\label{S_outer}\\
S_{\mathrm{back}}  &  \equiv\frac{1}{\sqrt{L}}\left(  \int_{0}^{rL^{1/2}}%
\frac{\exp(-r_{s}/l_{s})d\rho}{\left(  r^{2}-\rho^{2}\right)  ^{1/2}}%
+2\int_{rL^{1/2}}^{r}\frac{\exp(-r_{s}/l_{s})d\rho}{\left(  r^{2}-\rho
^{2}\right)  ^{1/2}}\right)  . \label{S_inner}%
\end{align}
The function $S_{\mathrm{for}}$ describes photons scattered forward
($\Theta_{\gamma}<90^{\circ}$). It includes only outgoing photons with
$r>r_{s}$ ($0<\rho<rL^{1/2}$). The function $S_{\mathrm{back}}$ describes
photons scattered backward ($\Theta_{\gamma}>90^{\circ}$). In the
\textquotedblleft inner\textquotedblright\ zone, $r<r_{s}$ ($rL^{1/2}<\rho
<r$), it includes both ingoing and outgoing photons with equal contributions,
while in the \textquotedblleft outer\textquotedblright\ zone, $r>r_{s}$, it
includes only outgoing photons. The integrals over $\rho$ describe the
original scattering of photons from the primary beams, where the distance of
scattering, $r_{s}=\rho/L^{1/2}$, depends on the angular parameter $\rho$ and
on the photon momenta via $L(p_{\gamma},p_{s})$. In the near zone,
$r/l_{s}\lesssim1$, all outer photons at given $r$ were scattered at distances
$r_{s}<r$. The situation for the inner photons is different. Some ingoing
photons at given $r$ could be originally backscattered at large distances,
even $r_{s}\gg r$ (for $L\ll1$), well beyond the near zone. The total
contribution of these photons, however, should be relatively small due to
exponential attenuation of the primary beams described by $\exp(-r_{s}%
/l_{s})=\exp[-\rho/(L^{1/2}l_{s})]$. Temporarily introducing%
\end{subequations}
\begin{equation}
P\left(  L,\lambda_{s}\right)  \equiv\int_{0}^{L^{1/2}}\frac{\exp\left(
-\lambda_{s}t\right)  dt}{\left(  1-t^{2}\right)  ^{1/2}}, \label{lambda_P}%
\end{equation}%
\begin{equation}
\lambda_{s}(r,p_{\gamma},p_{s})\equiv\frac{r}{L^{1/2}l_{s}}=\left(
\frac{m_{e}c}{p_{\gamma}}-\frac{m_{e}c}{p_{s}}\right)  ^{-1/2}\left(
\frac{m_{e}c}{p_{s}}-\frac{m_{e}c}{p_{\gamma}}+2\right)  ^{-1/2}\frac{r}%
{l_{s}}, \label{lambda}%
\end{equation}
and $t\equiv\rho/r$, we can rewrite the functions $S_{\mathrm{for}}$ and
$S_{\mathrm{back}}$ as%
\begin{equation}
S_{\mathrm{for}}=\frac{P\left(  L,\lambda_{s}\right)  }{\sqrt{L}},\qquad
S_{\mathrm{back}}=\frac{2P\left(  1,\lambda_{s}\right)  -P\left(
L,\lambda_{s}\right)  }{\sqrt{L}}. \label{PPP}%
\end{equation}
The transition from $S_{\mathrm{for}}$ to $S_{\mathrm{back}}$ at $L=1$
corresponding to $p_{\gamma}=p_{\mathrm{cr}s}$ is smooth. For example,
$P\left(  L,0\right)  =\arcsin(L^{1/2})$, while $2P\left(  1,0\right)
-P\left(  L,0\right)  =\pi-\arcsin(L^{1/2})$ is the continuation of the
multivalued function $\arcsin(L^{1/2})$ from its first quadrant ($0\leq
\arcsin(L^{1/2})<\pi/2$) to the second one ($\pi/2\leq\arcsin(L^{1/2})<\pi$).
A similar smooth transition takes place for non-zero $\lambda_{s}$.

If $\lambda_{s}\lesssim1$ then the scatter distances $r_{s}$ lie within the
near zone. However, as mentioned above, ingoing photons scattered backward can
arrive from arbitrarily large distances $r_{s}$, so we should consider any
possible values of $\lambda_{s}\geq0$. All such photons are described in
(\ref{PPP}) by the function $P\left(  1,\lambda_{s}\right)  $ shown in
Fig.\ \ref{Fig:P(1,lambda)_new}
%TCIMACRO{\FRAME{ftbpFU}{6.6701in}{5.1605in}{0pt}{\Qcb{Function $P(1,\lambda
%_{s})$ given by (\ref{P(1,lambda)}) (solid black), along with its
%approximation given by (\ref{integral_approx}) (dashed red).}}%
%{\Qlb{Fig:P(1,lambda)_new}}{Figure}{\special{ language "Scientific Word";
%type "GRAPHIC";  maintain-aspect-ratio TRUE;  display "USEDEF";
%valid_file "T";  width 6.6701in;  height 5.1605in;  depth 0pt;
%original-width 8.4618in;  original-height 6.5388in;  cropleft "0";
%tempfilename 'LXYN9F0K.wmf';tempfile-properties "XPR";}} }%
%BeginExpansion
\begin{figure}
[ptb]
\begin{center}
\includegraphics[
natheight=5.160500in,
natwidth=6.670100in,
height=5.1605in,
width=6.6701in
]%
{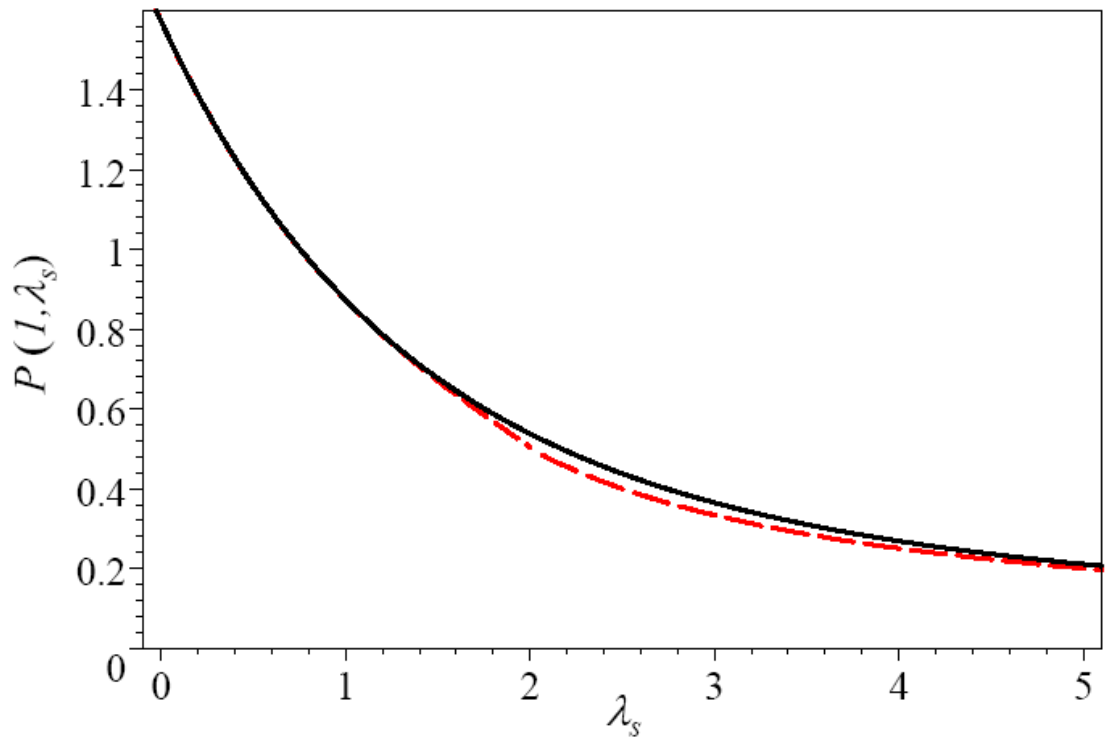}%
\caption{Function $P(1,\lambda_{s})$ given by (\ref{P(1,lambda)}) (solid
black), along with its approximation given by (\ref{integral_approx}) (dashed
red).}%
\label{Fig:P(1,lambda)_new}%
\end{center}
\end{figure}
%EndExpansion
and expressed analytically as
\begin{align}
&  P\left(  1,\lambda_{s}\right)  =\frac{\pi}{2}\left(  I_{0}(\lambda
_{s})-L_{0}(\lambda_{s})\right) \nonumber\\
&  =\frac{\pi}{2}\left[  1+\sum_{k=1}^{\infty}\frac{\lambda_{s}^{2k}}%
{2^{2k}k\left(  k-1\right)  !k!}\right]  -\lambda_{s}\left\{  1+\sum
_{k=1}^{\infty}\left[  \frac{2^{k}\left(  k!\right)  ^{k}\lambda_{s}^{k}%
}{\left(  2k+1\right)  !}\right]  ^{2}\right\}  , \label{P(1,lambda)}%
\end{align}
where $I_{0}(x)$ and $L_{0}(x)$ are the modified Bessel and Struve functions
\cite{AbramowitzStegun}. For a simpler analysis, one can approximate $P\left(
1,\lambda_{s}\right)  $ in terms of elementary functions by combining the
power series (\ref{P(1,lambda)}) to the fifth-order term with the
large-$\lambda_{s}$ asymptotics, $P\left(  1,\lambda_{s}\right)
\rightarrow1/\lambda_{s}$, as%
\begin{equation}
P\left(  1,\lambda_{s}\right)  \approx\left\{
\begin{array}
[c]{ccc}%
\frac{\pi}{2}\left(  1+\frac{\lambda_{s}^{2}}{8}\right)  ^{2}-\lambda
_{s}\left(  1+\frac{\lambda_{s}^{2}}{9}+\frac{\lambda_{s}^{4}}{225}\right)  &
\text{if} & \lambda_{s}\leq2\\
1/\lambda_{s} & \text{if} & \lambda_{s}>2
\end{array}
\right.  . \label{integral_approx}%
\end{equation}
This piece-wise approximation mimics the actual $P\left(  1,\lambda
_{s}\right)  $ almost ideally except for a minor kink at $1.5\lesssim
\lambda_{s}\lesssim4$, as seen in Fig.\ \ref{Fig:P(1,lambda)_new}.

Unlike $P\left(  1,\lambda_{s}\right)  $, the function $P\left(  L,\lambda
_{s}\right)  $ in (\ref{PPP}) describes only outgoing photons scattered at
distances $r_{s}$ closer to the source than the given distance $r$. According
to the near-zone restriction, $r/l_{s}\lesssim1$, in the entire integration
domain of (\ref{lambda_P}), $\lambda_{s}t\lesssim1$ holds. This allows us to
expand the exponential function in the integrand of $P\left(  L,\lambda
_{s}\right)  $ in powers of $(\lambda_{s}t)$ to the fifth power (applicable
for $\lambda_{s}t\leq2$). This yields
\begin{align}
P\left(  L,\lambda_{s}\right)   &  \approx\left(  1+\frac{\lambda_{s}^{2}}%
{8}\right)  ^{2}\arcsin\sqrt{L}-\lambda_{s}\left(  1+\frac{\lambda_{s}^{2}}%
{9}+\frac{\lambda_{s}^{4}}{225}\right)  \left(  1-\sqrt{1-L}\right)
\nonumber\\
&  -\frac{\lambda_{s}^{2}\sqrt{L\left(  1-L\right)  }}{4}\left\{
1+\frac{\lambda_{s}^{2}}{16}-\frac{\lambda_{s}\sqrt{L}}{9}\left[
2+\frac{\lambda_{s}^{2}\left(  4+3L\right)  }{50}-\frac{3\lambda_{s}\sqrt{L}%
}{8}\right]  \right\}  . \label{P_approx_5}%
\end{align}%
%TCIMACRO{\FRAME{ftbpFU}{5.0496in}{3.7972in}{0pt}{\Qcb{Photon energy
%distribution for unshielded $_{27}^{60}$Co at various distances $r$. The
%analytical solutions are shown by solid lines.}}{\Qlb{Fig:Energy_distribution}%
%}{Figure}{\special{ language "Scientific Word";  type "GRAPHIC";
%maintain-aspect-ratio TRUE;  display "USEDEF";  valid_file "T";
%width 5.0496in;  height 3.7972in;  depth 0pt;  original-width 5.1188in;
%original-height 3.8415in;  cropleft "0";  croptop "1";  cropright "1";
%cropbottom "0";  tempfilename 'LY1RMY00.wmf';tempfile-properties "XPR";}} }%
%BeginExpansion
\begin{figure}
[ptb]
\begin{center}
\includegraphics[
natheight=3.797200in,
natwidth=5.049600in,
height=3.7972in,
width=5.0496in
]%
{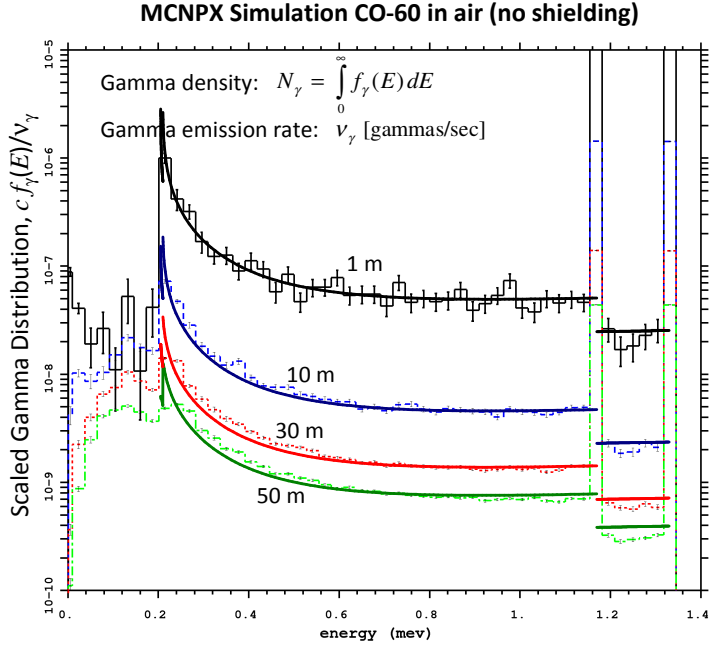}%
\caption{Photon energy distribution for unshielded $_{27}^{60}$Co at various
distances $r$. The analytical solutions are shown by solid lines.}%
\label{Fig:Energy_distribution}%
\end{center}
\end{figure}
%EndExpansion
%TCIMACRO{\FRAME{ftbpFU}{5.0496in}{3.7972in}{0pt}{\Qcb{Photon energy
%distribution corresponding to Fig.\ \ref{Fig:Energy_distribution}, but for the
%1 cm thick spherical lead shielding placed at 99 cm from the source.}%
%}{\Qlb{Fig:Shielded}}{Figure}{\special{ language "Scientific Word";
%type "GRAPHIC";  maintain-aspect-ratio TRUE;  display "USEDEF";
%valid_file "T";  width 5.0496in;  height 3.7972in;  depth 0pt;
%original-width 5.1188in;  original-height 3.8415in;  cropleft "0";
%tempfilename 'LY1RMZ01.wmf';tempfile-properties "XPR";}} }%
%BeginExpansion
\begin{figure}
[ptb]
\begin{center}
\includegraphics[
natheight=3.797200in,
natwidth=5.049600in,
height=3.7972in,
width=5.0496in
]%
{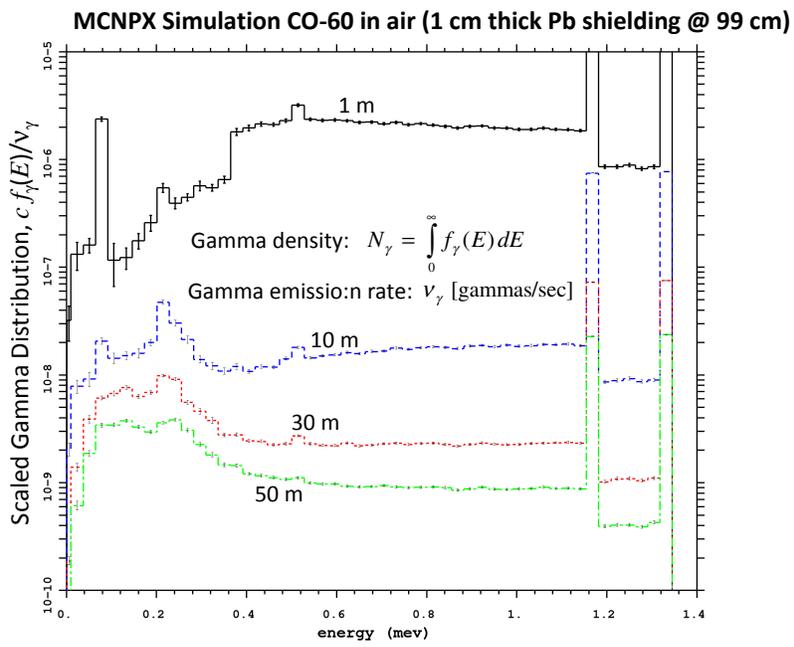}%
\caption{Photon energy distribution corresponding to
Fig.\ \ref{Fig:Energy_distribution}, but for the 1 cm thick spherical lead
shielding placed at 99 cm from the source.}%
\label{Fig:Shielded}%
\end{center}
\end{figure}
%EndExpansion

Returning from the temporary notation $\lambda_{s}=r/(l_{s}\sqrt{L})$
(\ref{lambda}) back to the distance $r$, we obtain
\begin{align}
&  S_{\mathrm{for}}=\left(  1+\frac{r^{2}}{8l_{s}^{2}L}\right)  ^{2}%
\frac{\arcsin\sqrt{L}}{\sqrt{L}}-\frac{r\left(  1-\sqrt{1-L}\right)  }{l_{s}%
L}\left(  1+\frac{r^{2}}{9l_{s}^{2}L}+\frac{r^{4}}{225l_{s}^{4}L^{2}}\right)
\nonumber\\
&  -\frac{r^{2}\sqrt{1-L}}{4l_{s}^{2}L}\left\{  1+\frac{r^{2}}{16l_{s}^{2}%
L}-\frac{r}{9l_{s}}\left[  2+\frac{r^{2}\left(  4+3L\right)  }{50l_{s}^{2}%
L}-\frac{3r}{8l_{s}}\right]  \right\}  , \label{P_L}%
\end{align}%
\begin{equation}
S_{\mathrm{back}}=\frac{\pi}{\sqrt{L}}\left[  I_{0}\left(  \frac{r}{l_{s}%
\sqrt{L}}\right)  -L_{0}\left(  \frac{r}{l_{s}\sqrt{L}}\right)  \right]
-S_{\mathrm{forward}}, \label{P-P_L}%
\end{equation}
where, according to (\ref{P(1,lambda)})--(\ref{integral_approx}), the first
term in the RHS of (\ref{P-P_L}) can be approximated by elementary functions
as%
\begin{align}
&  \frac{\pi}{\sqrt{L}}\left[  I_{0}\left(  \frac{r}{l_{s}\sqrt{L}}\right)
-L_{0}\left(  \frac{r}{l_{s}\sqrt{L}}\right)  \right] \nonumber\\
&  \approx\left\{
\begin{array}
[c]{ccc}%
\frac{\pi}{\sqrt{L}}\left(  1+\frac{r^{2}}{8l_{s}^{2}L}\right)  ^{2}-\frac
{2r}{l_{s}L}\left(  1+\frac{r^{2}}{9l_{s}^{2}L}+\frac{r^{4}}{225l_{s}^{4}%
L^{2}}\right)  & \text{if} & \sqrt{L}\geq r/(2l_{s})\\
2l_{s}/r & \text{if} & \sqrt{L}<r/(2l_{s})
\end{array}
\right.  . \label{approxa}%
\end{align}

The function $S_{\mathrm{for}}$ describes photons scattered forward
($0^{\circ}<\Theta_{\gamma}\leq90^{\circ}$, $p_{\mathrm{cr}s}\leq p_{\gamma
}<p_{\max}=p_{s}$), while $S_{\mathrm{back}}$ describes photons scattered
backward ($90^{\circ}<\Theta_{\gamma}\leq180^{\circ}$, $p_{\min}%
=p_{s}/[1+2p_{s}/(m_{e}c)]<p_{\gamma}<p_{\mathrm{cr}s}$). The two functions
are smoothly connected at $L=1$ ($\Theta_{\gamma}=90^{\circ}$, $p_{\gamma
}=p_{\mathrm{cr}s}$). In spite of singular denominators $L^{-q}$ in
(\ref{P_L})--(\ref{approxa}), $S_{\mathrm{for}}$ and $S_{\mathrm{back}}$
behave regularly at $L=0$ ($\Theta_{\gamma}=0^{\circ},180^{\circ}$), except
for $r=0$. At $r\ll l_{s}$, $S_{\mathrm{for}}$ ($\Theta_{\gamma}=0^{\circ}$,
$p_{\gamma}=p_{s}$) approaches $1$, while $S_{\mathrm{back}}$ ($\Theta
_{\gamma}=180^{\circ}$, $p_{\gamma}=p_{\min}$) increases with decreasing $r$
as $2l_{s}/r$. This $1/r$ divergence of $S_{\mathrm{back}}(L=0)$ at
$r\rightarrow0$ is due to accumulation of photons scattered almost strictly
backward from distances $r_{s}\gg r$ where the primary beams still remain
largely unattenuated, $r_{s}\ll l_{s}$. As $r$ increases, even remaining
within the mean free path $l_{s}$ from the radiation source, the exponential
attenuation of the primary beams at larger distances $r_{s}\gtrsim l_{s}$
makes such accumulation less significant.

Inserting (\ref{P_L}) and (\ref{P-P_L}) into (\ref{F_prome}), with $L$ given
in terms of $p_{\gamma},p_{s}$ by (\ref{sin^2Theta}), i.e.,%
\begin{align}
\sqrt{L}  &  =\sin\Theta_{\gamma}=\left(  \frac{m_{e}c}{p_{\gamma}}%
-\frac{m_{e}c}{p_{s}}\right)  ^{1/2}\left(  \frac{m_{e}c}{p_{s}}-\frac{m_{e}%
c}{p_{\gamma}}+2\right)  ^{1/2},\nonumber\\
\sqrt{1-L}  &  =\left\vert \cos\Theta_{\gamma}\right\vert =\left\vert
1-\frac{m_{e}c}{p_{\gamma}}+\frac{m_{e}c}{p_{s}}\right\vert , \label{prohe_L}%
\end{align}
we obtain the complete approximate analytical expression for $F(p_{\gamma})$
(\ref{summa_F}) in the near zone.

Figure\ \ref{Fig:Energy_distribution} shows photon energy distributions for
unshielded $_{27}^{60}$Co calculated by the MCNPX code (URL: http://mcnpx.lanl.gov/) and the analytical solution given by (\ref{F_prome}) and (\ref{P_L})--(\ref{prohe_L}) for several
distances $r$ (shown near the curves). The two long spikes on the right
correspond to the two monochromatic primary photon beams with $\mathcal{E}%
_{1}=1.332$ and $\mathcal{E}_{2}=1.173$~MeV, described analytically by
(\ref{F_prim}). We see an excellent agreement between the simulations and
theory in the entire energy range of singly scattered photons, $\mathcal{E}%
_{2}/\left(  1+2\mathcal{E}_{2}/m_{e}\right)  \approx0.21\
%TCIMACRO{\unit{MeV}}%
%BeginExpansion
\operatorname{MeV}%
%EndExpansion
<\mathcal{E}<1.332\
%TCIMACRO{\unit{MeV}}%
%BeginExpansion
\operatorname{MeV}%
%EndExpansion
$. At $\mathcal{E}<0.21\
%TCIMACRO{\unit{MeV}}%
%BeginExpansion
\operatorname{MeV}%
%EndExpansion
$, the simulations show lower-energy multiply scattered photons excluded from
the extended $n=1$ model. In the near zone, especially due to the weighting
factor $p_{\gamma}$ in the integrand of (\ref{dN_e/dt_via_p}), the
lower-energy range makes a small relative contribution to $dN_{e}/dt$.

Figure\ \ref{Fig:Shielded} shows simulated photon energy distributions for the
same radioactive source but shielded with a thin spherical lead cover. The
clearly pronounced broad bumps at the high-energy tails of the photon
distribution, especially at the closest proximity to the source ($r=1\
%TCIMACRO{\unit{m}}%
%BeginExpansion
\operatorname{m}%
%EndExpansion
$) are due to primary photons scattered within the shield material and leaked
outside. This scattering could also be described by the general solution
(\ref{F_prim})--(\ref{recur_relation}) with $\hat{L}_{r}$ defined by
(\ref{I_arr}), where the collisional mean free path for photons in the
shielding metal is many orders of magnitude smaller than that in air.
Unfortunately, for dense shielding the near-zone approximation is hardly
applicable. Besides, photoemission absorption may play a role. This makes
analytical calculation much more difficult. Comparing
Figs.\ \ref{Fig:Energy_distribution} and \ref{Fig:Shielded}, we see that the
effect of shielding on the energy distribution is especially pronounced in a
close proximity to the source. For distances $\gtrsim50$\ m, the photon energy
distribution in both shielded and unshielded cases look similar.

\subsection{Electron production rate}

Now we calculate the electron production rate (\ref{dN_e/dt_via_p}) due to
scattered photons with the energy distribution function given by
(\ref{F_prome}), or%
\[
F_{s}(p_{\gamma}=m_{e}c\alpha)=\frac{Z_{\mathrm{air}}I_{\gamma}^{(0)}%
n_{m}r_{0}^{2}}{2rm_{e}c^{2}}\frac{1}{\alpha_{s}^{2}}\left(  \frac{\alpha
}{\alpha_{s}}+\frac{\alpha_{s}}{\alpha}-L\right)  \left\{
\begin{array}
[c]{ccc}%
S_{\mathrm{for}} & \text{if} & \alpha\geq\frac{\alpha_{s}}{1+\alpha_{s}}\\
S_{\mathrm{back}} & \text{if} & \alpha<\frac{\alpha_{s}}{1+\alpha_{s}}%
\end{array}
\right.  ,
\]
where $\alpha_{s}\equiv p_{s}/(m_{e}c)$, $L=1-\left(  1-1/\alpha+1/\alpha
_{s}\right)  ^{2}$ and $S_{\mathrm{for}}$, $S_{\mathrm{back}}$ are given by
(\ref{P_L})--(\ref{approxa}). We restrict calculations to the radiation
material $_{27}^{60}$Co with $E_{1}=1.332\
%TCIMACRO{\unit{MeV}}%
%BeginExpansion
\operatorname{MeV}%
%EndExpansion
$ and $E_{2}=1.173\
%TCIMACRO{\unit{MeV}}%
%BeginExpansion
\operatorname{MeV}%
%EndExpansion
$.

\begin{figure}
[ptb]
\begin{center}
\includegraphics[
natheight=5.157800in,
natwidth=6.674500in,
height=5.1578in,
width=6.6745in
]%
{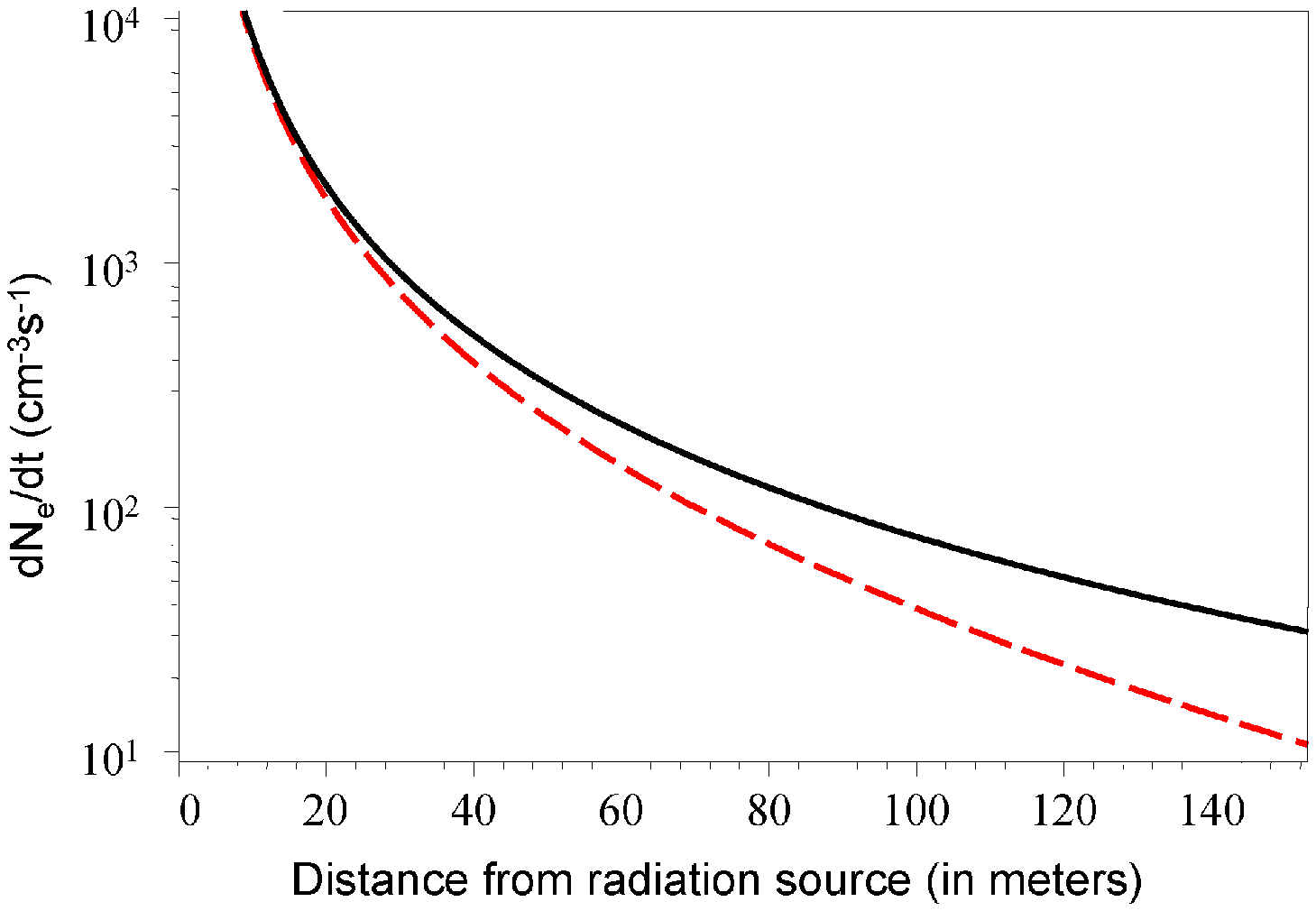}%
\caption{Free-electron production rate for $1\operatorname{Ci}$
($0.\allowbreak9\operatorname{mg}$) of $_{27}^{60}$Co as a function of the
distance from the raioactive source. The red dashed line shows the
primary-photon rate given by (\ref{dN_e/dt_prim}), while he black sold line
shows the total rate given by (\ref{dN_e/dt_total}).}%
\label{Fig:dN/dt}%
\end{center}
\end{figure}

When calclulating the production rate for scattered photons, $(dN_{e}%
/dt)_{\mathrm{scat}}$, it is convenient to compare it with production rate for
primary photons given by (\ref{dN_e/dt_prim}). Straightforward numerical
integration shows that for each $s=1,2$ a simple approximation%
\begin{align}
&  \frac{1}{\alpha_{s}^{2}}\left[  \int_{\frac{\alpha_{s}}{1+2\alpha_{s}}%
}^{\frac{\alpha_{s}}{1+\alpha_{s}}}\left(  \frac{\alpha}{\alpha_{s}}%
+\frac{\alpha_{s}}{\alpha}-L\right)  S_{\mathrm{back}}\alpha d\alpha
+\int_{\frac{\alpha_{s}}{1+\alpha_{s}}}^{\alpha_{s}}\left(  \frac{\alpha
}{\alpha_{s}}+\frac{\alpha_{s}}{\alpha}-L\right)  S_{\mathrm{for}}\alpha
d\alpha\right]  \nonumber\\
&  \simeq1.37\exp\left(  -0.42r/l_{s}\right)  \label{1.37exp()}%
\end{align}
holds to a better than 5\% accuracy. To compare with $(dN_{e}%
/dt)_{\mathrm{prim}}$, we express $l_{s}=c/\nu_{s}$, with $\nu_{s}$ in terms
of $\alpha_{s}$ by (\ref{nu_KN(p)}), and obtain%
\begin{align*}
&  \left.  \frac{Z_{\mathrm{air}}I_{\gamma}^{(0)}n_{m}r_{0}^{2}}{2r\alpha
_{s}^{2}m_{e}c^{2}}\right/  \frac{I_{\gamma}^{(0)}}{4\pi cr^{2}p_{s}}\\
&  =\frac{r}{2l_{s}}\left[  \frac{2+8\alpha_{s}+9\alpha_{s}^{2}+\alpha_{s}%
^{3}}{\alpha_{s}\left(  1+2\alpha_{s}\right)  ^{2}}-\frac{\left(
2+2\alpha_{s}-\alpha_{s}^{2}\right)  \ln\left(  1+2\alpha_{s}\right)
}{2\alpha_{s}^{2}}\right]  ^{-1}.
\end{align*}
As a result, the total electron production rate becomes%
\begin{equation}
\left(  \frac{dN_{e}}{dt}\right)  _{\mathrm{rad}}=3.5\times10^{5}%
%TCIMACRO{\unit{cm}}%
%BeginExpansion
\operatorname{cm}%
%EndExpansion
^{-3}%
%TCIMACRO{\unit{s}}%
%BeginExpansion
\operatorname{s}%
%EndExpansion
^{-1}\ \frac{I_{\gamma}^{(0)}(%
%TCIMACRO{\unit{Ci}}%
%BeginExpansion
\operatorname{Ci}%
%EndExpansion
)}{r^{2}(%
%TCIMACRO{\unit{m}}%
%BeginExpansion
\operatorname{m}%
%EndExpansion
)}\sum_{s=1}^{k}\mathcal{E}_{s}(%
%TCIMACRO{\unit{MeV}}%
%BeginExpansion
\operatorname{MeV}%
%EndExpansion
)\left(  e^{-r/l_{s}}+\frac{\mu_{s}r}{l_{s}}\ e^{-0.42r/l_{s}}\right)
\label{dN_e/dt_total}%
\end{equation}
where the last term in the RHS brackets represents the ratio between the
production rate due to the scattered photons to that due to the primary
photons (for each primary beam $s$) and%
\begin{equation}
\mu_{s}=\frac{0.69}{\alpha_{s}\left[  \frac{1+\alpha_{s}}{\alpha_{s}^{2}%
}\left(  \frac{2\left(  1+\alpha_{s}\right)  }{1+2\alpha_{s}}-\frac{\ln\left(
1+2\alpha_{s}\right)  }{\alpha_{s}}\right)  +\frac{\ln\left(  1+2\alpha
_{s}\right)  }{2\alpha_{s}}-\frac{1+3\alpha_{s}}{\left(  1+2\alpha_{s}\right)
^{2}}\right]  }\label{mu_s}%
\end{equation}
Equation (\ref{dN_e/dt_total}) shows that the scattered-photon contribition to
the total free-electron production rate decays with distance as
$e^{-0.42r/l_{s}}/r$, i.e. noticeably slower than that of the primary photons,
$e^{-r/l_{s}}/r^{2}$. For $_{27}^{60}$Co, we have $l_{1}\approx127$\ $%
%TCIMACRO{\unit{m}}%
%BeginExpansion
\operatorname{m}%
%EndExpansion
$, $\alpha_{s}\approx2.607$, $\mu_{1}\approx$ $0.72$, and $l_{2}\simeq119$\ $%
%TCIMACRO{\unit{m}}%
%BeginExpansion
\operatorname{m}%
%EndExpansion
$, $\alpha_{2}\approx2.295$, $\mu_{2}\simeq0.77$. The corresponding $\left(
dN_{e}/dt\right)  _{\mathrm{rad}}$ is shown in Fig.~\ref{Fig:dN/dt}, along
with the primary photon rate given by (\ref{dN_e/dt_prim}). This Figure shows
that the latter is a good approximation for $r<40%
%TCIMACRO{\unit{m}}%
%BeginExpansion
\operatorname{m}%
%EndExpansion
$. At larger distances, $dN_{e}/dt$ due to scattered photons becomes
noticeable and dominates starting from $r\simeq100\
%TCIMACRO{\unit{m}}%
%BeginExpansion
\operatorname{m}%
%EndExpansion
$. For simple estimates, we can extrapolate equation (\ref{dN_e/dt_total}) to
larger distances, but for our application only moderate distances, $r<150\
%TCIMACRO{\unit{m}}%
%BeginExpansion
\operatorname{m}%
%EndExpansion
$, matter (see below).

Figure \ref{Fig:dN/dt} also shows that a moderate amount ($\sim1\
%TCIMACRO{\unit{mg}}%
%BeginExpansion
\operatorname{mg}%
%EndExpansion
$) of a concealed radioactive material like $_{27}^{60}$Co can provide the
total rate of electron production well above the background level
($Q_{\mathrm{rad}}\sim20$ pairs $%
%TCIMACRO{\unit{s}}%
%BeginExpansion
\operatorname{s}%
%EndExpansion
^{-1}%
%TCIMACRO{\unit{cm}}%
%BeginExpansion
\operatorname{cm}%
%EndExpansion
^{-3}$). The enhancement factor $\alpha_{\mathrm{rad}}=\left(  dN_{e}%
/dt\right)  _{\mathrm{rad}}/Q_{\mathrm{rad}}$ exceeds unity within $150\
%TCIMACRO{\unit{m}}%
%BeginExpansion
\operatorname{m}%
%EndExpansion
$ from the radiation source. This distance is crucial for implementation of
our technique because ... As discussed in Ref.~\cite{NusinSprangle}, all free
electrons will eventually end up forming negative ions, but the collisional
detachment of some electrons for a brief time make it sufficient to initiate
air breakdown within the focal spot of the THz gyrotron during pulse.
Equation~(\ref{dN_e/dt_total}) provides a reliable estimate of the electron
production rate at least within the distance of $150\
%TCIMACRO{\unit{m}}%
%BeginExpansion
\operatorname{m}%
%EndExpansion
$ critical for implementation of our method.

\paragraph{Conclusion:}

In this paper, we analyze the production of free electrons by $\gamma$-rays
leaking from radioactive materials. This is important for implementation of a
recently proposed concept to remotely detect concealed radioactive materials
\cite{Gr.Nusin.2010}, \cite{NusinSprangle}. We have performed our analysis
specifically for the most probable radioactive material $_{27}^{60}$Co, but
the general approach can be applied to any $\gamma$-ray sources. In our
theory, we have included free-electron production by both primary $\gamma
$-quanta radiated by the source and penetrated through the container walls and
those scattered in air. Equation (\ref{dN_e/dt_total}) and
Fig.~\ref{Fig:dN/dt} give a reliable estimate of the total production rate
$\left(  dN_{e}/dt\right)  _{\mathrm{rad}}/Q_{\mathrm{rad}}$ and the
corresponding radiation enhancement factor, at least within $150$ $%
%TCIMACRO{\unit{m}}%
%BeginExpansion
\operatorname{m}%
%EndExpansion
$ from the radiation source. At these distances, the scattered photon
contribution to $\left(  dN_{e}/dt\right)  _{\mathrm{rad}}$ is comparable or
even exceeds that of the primary photons. Even for amount of radiation
material as small as $\sim1\
%TCIMACRO{\unit{mg}}%
%BeginExpansion
\operatorname{mg}%
%EndExpansion
$, the total production rate may exceed significantly the ambient ionization
rate. During the gyrotron pulse of about 10 microsecond length, such electrons
may seed the electric breakdown and create sufficiently dense plasma at the
focal region to be detected as an unambiguous effect of the concealed
radioactive material.%
%TCIMACRO{\FRAME{ftbpFU}{6.6745in}{5.1578in}{0pt}{\Qcb{Free-electron production
%rate for $1\unit{Ci}$ ($0.\allowbreak9\unit{mg}$) of $_{27}^{60}$Co as a
%function of the distance from the raioactive source. The red dashed line shows
%the primary-photon rate given by (\ref{dN_e/dt_prim}), while he black sold
%line shows the total rate given by (\ref{dN_e/dt_total}).}}{\Qlb{Fig:dN/dt}%
%}{dnedt.tif}{\special{ language "Scientific Word";  type "GRAPHIC";
%width 6.6745in;  height 5.1578in;  depth 0pt;  original-width 11.2611in;
%original-height 8.6924in;  cropleft "0";  croptop "1";  cropright "1";
%cropbottom "0";  filename 'dNedt.tif';file-properties "XNPEU";}} }%
%BeginExpansion
%EndExpansion

\paragraph{Acknowledgement:}

This work is supported by the US Office of Naval Research.

\bigskip\bigskip

\appendix\noindent{\textbf{{\Large Appendix}}}

\setcounter{equation}{0} \renewcommand{\theequation}{\thesection.\arabic{equation}}

\section{Deriving kinetic equation (\ref{kinEq})
\label{Deriving kinetic equation}}

The Boltzmann kinetic equation for freely propagating unpolarized photons
colliding with air molecules contains two parts: the collisionless part
describing the undisturbed propagation of photons between two consecutive acts
of scattering or absorption and the collisional operator \cite{LandauKinetics}%
. In the assumed spherically symmetric approximation, the stationary kinetic
equation for $f_{\gamma}^{\pm}\left(  r,p,\rho\right)  $ with $\rho$ given by
(\ref{rho}) can be written as
\begin{equation}
\pm c\left(  1-\frac{\rho^{2}}{r^{2}}\right)  ^{1/2}\frac{\partial f_{\gamma
}^{\pm}\left(  r,p,\rho\right)  }{\partial r}=\left[  \frac{df_{\gamma}^{\pm}%
}{dt}\right]  _{\mathrm{coll}}. \label{stac_kin}%
\end{equation}
The LHS of (\ref{stac_kin}) contains no $p,\rho$-derivatives because these
variables are invariants of the collisionless motion. The collisional term
$\left[  df_{\gamma}^{\pm}/dt\right]  _{\mathrm{coll}}$ describes scattering,
absorption and other possible effects responsible for collisional changes in
the photon momentum distribution. In the general case, it has a complicated
integral form which can be simplified in the spherically symmetric case.

In this appendix, we obtain the explicit simplified form for $\left[
df_{\gamma}^{\pm}/dt\right]  _{\mathrm{coll}}$. We will start directly from
the original probabilistic expressions like \cite{LandauKinetics}. Since the
photon-molecule collision is a strictly local process where the photon
transport plays no role, in the RHS\ of (\ref{stac_kin})\ we can temporarily
return from $\rho$ back to the momentum polar angle $\theta_{\gamma}$.
\begin{align}
&  \left(  \frac{df_{\gamma}}{dt}\right)  _{\mathrm{coll}}=I_{\mathrm{arr}%
}-I_{\mathrm{dep}},\nonumber\\
I_{\mathrm{arr}}  &  =\int f_{\gamma}^{\prime}(r,\vec{p}_{\gamma}^{\prime
})W\left(  \vec{p}_{\gamma}^{\prime}\rightarrow\vec{p}_{\gamma}\right)
\nonumber\\
&  \times\delta(p_{\gamma}+\frac{\mathcal{E}_{e}}{c}-p_{\gamma}^{\prime}%
-m_{e}^{\prime}c)\delta(\vec{p}_{\gamma}+\vec{p}_{e}-\vec{p}_{\gamma}^{\prime
})d^{3}p_{\gamma}^{\prime}d^{3}p_{e},\nonumber\\
I_{\mathrm{dep}}  &  =f_{\gamma}(r,\vec{p}_{\gamma})\int W\left(  \vec
{p}_{\gamma}\rightarrow\vec{p}_{\gamma}^{\prime}\right)  \label{coll_operator}%
\\
&  \times\delta(p_{\gamma}^{\prime}+\frac{\mathcal{E}_{e}^{\prime}}%
{c}-p_{\gamma}-m_{e}c)\delta(\vec{p}_{\gamma}^{\prime}+\vec{p}_{e}^{\prime
}-\vec{p}_{\gamma})d^{3}p_{\gamma}^{\prime}d^{3}p_{e}^{\prime}.\nonumber
\end{align}
Here $I_{\mathrm{dep}}$ describes the total collisional departure of photons
from the kinetic volume $d^{3}p_{\gamma}$ around given $\vec{p}_{\gamma}%
$\ with scattered photon momenta in the domain $d^{3}p_{\gamma}^{\prime}$ and
releasing electrons in $d^{3}p_{e}^{\prime}$ around $\vec{p}_{\gamma}^{\prime
}$ and $\vec{p}_{e}^{\prime}$, respectively. In the general case, the
departure term, $I_{\mathrm{dep}}$, can include both photon scattering and
absorption. The term $I_{\mathrm{arr}}$ describes the arrival of photons and
released electrons into the volumes $d^{3}p_{\gamma}$ and $d^{3}p_{e}$ around
$\vec{p}_{\gamma}$ and $\vec{p}_{e}$, respectively, if the incident photons
before scattering were at $d^{3}p_{\gamma}^{\prime}$ around $\vec{p}_{\gamma
}^{\prime}$. Since we neglect any collisional photon production like
Bremsstrahlung, fluorescence, etc., $I_{\mathrm{arr}}$ includes only photon
scattering. With no photon losses included in $I_{\mathrm{dep}}$,
(\ref{coll_operator}) provides automatically the particle conservation,
$\int(df_{\gamma}/dt)_{\mathrm{coll}}d^{3}p_{\gamma}=0$.

The function $W$ of two arguments $\vec{p}_{\gamma}^{\prime}$ and $\vec
{p}_{\gamma}$ represents the probability per unit time of the collisional
process. This function is proportional to the air molecular density,
$n_{m}=n_{a}/2$. In the general case, the two arguments of $W$ are not
interchangeable. To identify the correct order of the arguments, the time
sequence of the collision is denoted by the arrow. The two delta-functions in
the integrands describe the energy and momentum conservation. Since we assume
energies well above the K-shell edge, the arguments of the delta functions
include no small energy spent on the molecule excitation and ionization. They
also neglect the kinetic energy and momentum of air molecules. As explained in
section\ \ref{Photon_Kinetics}, we presume Compton scattering in the K-N approximation.

To obtain the explicit expression for $W$, we relate it to the corresponding
differential collision cross section, $d\sigma/d\Omega_{\gamma}^{\prime}$,
given for the K-N theory by Eq.\ (\ref{Klein-Nishina}). To find the expression
is easier through the departure term $I_{\mathrm{dep}}$ because, unlike
$I_{\mathrm{arr}}$, it does not involve the photon distribution function in
the integrand. Integrating (\ref{coll_operator}), first, over $\vec{p}%
_{e}^{\prime}$ with eliminating $\delta(\vec{p}_{\gamma}^{\prime}+\vec{p}%
_{e}^{\prime}-\vec{p}_{\gamma})$ and then over $p_{\gamma}^{\prime}$ in
$d^{3}p_{\gamma}^{\prime}=p_{\gamma}^{\prime2}dp_{\gamma}^{\prime}%
d\Omega_{\gamma}^{\prime}$ with eliminating the remaining delta-function, we
obtain
\begin{equation}
2cZ_{\mathrm{air}}n_{m}\ \frac{d\sigma_{\mathrm{KN}}}{d\Omega_{\gamma}%
^{\prime}}=\left.  W\left(  \vec{p}_{\gamma}\rightarrow\vec{p}_{\gamma
}^{\prime}\right)  \right\vert _{\mathcal{E}_{e}^{\prime}/c=p_{\gamma}%
+m_{e}c-p_{\gamma}^{\prime}}\left(  1+\frac{1}{c}\frac{\partial\mathcal{E}%
_{e}^{\prime}}{\partial p_{\gamma}^{\prime}}\right)  ^{-1}p_{\gamma}^{\prime
2}. \label{rela_1}%
\end{equation}
Here $2Z_{\mathrm{air}}n_{m}$ is the total density of bound electrons and
$\partial\mathcal{E}_{e}^{\prime}/\partial p_{\gamma}^{\prime}$ should be
calculated from the momentum conservation, $\vec{p}_{\gamma}=\vec{p}_{\gamma
}^{\prime}+\vec{p}_{e}^{\prime}$, but before applying the energy conservation,
$p_{\gamma}+m_{e}c=\mathcal{E}_{e}^{\prime}/c+p_{\gamma}^{\prime}$. As a
result, we arrive at%
\begin{equation}
W\left(  \vec{p}_{\gamma}\rightarrow\vec{p}_{\gamma}^{\prime}\right)
\overset{\mathrm{dep}}{=}\frac{Z_{\mathrm{air}}n_{m}r_{0}^{2}m_{e}c^{3}%
}{\mathcal{E}_{e}^{\prime}p_{\gamma}p_{\gamma}^{\prime}}\left(  \frac
{p_{\gamma}}{p_{\gamma}^{\prime}}+\frac{p_{\gamma}^{\prime}}{p_{\gamma}%
}-1+\cos^{2}\Theta_{\gamma}\right)  , \label{W_1}%
\end{equation}
where $\Theta_{\gamma}$ is the scatter angle from $\vec{p}_{\gamma}$\ to
$\vec{p}_{\gamma}^{\prime}$ given by (\ref{cos_theta_gamma}). This results in
$I_{\mathrm{dep}}=\nu(p_{\gamma})f_{\gamma}(r,\vec{p}_{\gamma})$, where
$\nu(p_{\gamma})$ is given by (\ref{nu_KN(p)}).

Now we proceed to calculating the more complicated arrival term,
$I_{\mathrm{arr}}$. Switching in (\ref{W_1}) the arguments of $W$, we obtain%
\begin{equation}
W\left(  \vec{p}_{\gamma}^{\prime}\rightarrow\vec{p}_{\gamma}\right)
\overset{\mathrm{arr}}{=}\frac{Z_{\mathrm{air}}n_{m}r_{0}^{2}m_{e}c^{3}%
}{\mathcal{E}_{e}p_{\gamma}p_{\gamma}^{\prime}}\left(  \frac{p_{\gamma}%
}{p_{\gamma}^{\prime}}+\frac{p_{\gamma}^{\prime}}{p_{\gamma}}-1+\cos^{2}%
\Theta_{\gamma}\right)  , \label{W_2}%
\end{equation}
where $\cos\Theta_{\gamma}$ in terms of $p_{\gamma}$ and $p_{\gamma}^{\prime
}\geq p_{\gamma}$ is given by (\ref{cos_Theta}). After inserting (\ref{W_2})
to (\ref{coll_operator}), as above, the first step to calculating
$I_{\mathrm{arr}}$ is to integrate over $\vec{p}_{e}$ with eliminating
$\delta(\vec{p}_{\gamma}+\vec{p}_{e}-\vec{p}_{\gamma}^{\prime})$. This yields
\begin{equation}
E_{e}^{2}\equiv p_{e}^{2}+m_{e}^{2}=p_{\gamma}^{2}+p_{\gamma}^{\prime2}%
+m_{e}^{2}-2p_{\gamma}p_{\gamma}^{\prime}\cos\Theta_{\gamma} \label{E_e_prom}%
\end{equation}
Unlike $I_{\mathrm{dep}}$, however, the arrival term $I_{\mathrm{arr}}$
contains the distribution function $f_{\gamma}^{\prime}\left(  r,\vec
{p}_{\gamma}^{\prime}\right)  $ within the integrand. This does not allow us
to easily integrate over $dp_{\gamma}^{\prime}$ in $d^{3}p_{\gamma}^{\prime
}=p_{\gamma}^{\prime2}dp_{\gamma}^{\prime}d\Omega_{\gamma}^{\prime}$ in order
to eliminate the remaining delta function. However, the fact that the
distribution function $f_{\gamma}^{\prime}$ depends on $p_{\gamma}^{\prime}$
and $\theta_{\gamma}^{\prime}$, rather than on the axial angle $\phi_{\gamma
}^{\prime}$ suggests eliminating the remaining delta function via the
integration over $\phi_{\gamma}^{\prime}$, using a simple geometric relation
\begin{equation}
\cos\Theta_{\gamma}=\cos\theta_{\gamma}^{\prime}\cos\theta_{\gamma}+\sin
\theta_{\gamma}^{\prime}\sin\theta_{\gamma}\cos\left(  \phi_{\gamma}^{\prime
}-\phi_{\gamma}\right)  . \label{angle_relation}%
\end{equation}
Given $p_{\gamma},\theta_{\gamma},p_{\gamma}^{\prime},\theta_{\gamma}^{\prime
}$, this yields $E_{e}^{2}\left(  \phi_{\gamma}^{\prime}\right)  $.
Integration of in (\ref{coll_operator}) over $\phi_{\gamma}^{\prime}$ using%
\begin{equation}
\delta(p_{\gamma}+E_{e}-p_{\gamma}^{\prime}-m_{e}^{\prime})=\left\vert
\frac{\partial E_{e}^{\prime}}{\partial\phi_{\gamma}^{\prime}}\right\vert
^{-1}\delta(\phi_{\gamma}^{\prime}-\phi_{\mathrm{c}}^{\prime})
\label{delta_prom}%
\end{equation}
results in
\begin{equation}
I_{\mathrm{arr}}=2\int f_{\gamma}^{\prime}W\left(  \vec{p}_{\gamma}^{\prime
}\rightarrow\vec{p}_{\gamma}\right)  \left\vert \frac{\partial E_{e}}%
{\partial\phi_{\gamma}^{\prime}}\right\vert ^{-1}p_{\gamma}^{\prime
2}dp_{\gamma}^{\prime}d\left(  \cos\theta_{\gamma}^{\prime}\right)  .
\label{prom}%
\end{equation}
The factor $2$ in (\ref{prom}) originates from existence of two symmetric
values of $\phi_{\gamma}^{\prime}=\phi_{\mathrm{c}}^{\prime}$ that set the
argument of the delta-function to zero (the energy conservation)
\begin{equation}
E_{e}=p_{\gamma}^{\prime}+m_{e}-p_{\gamma}. \label{E_e_final}%
\end{equation}
The specific value of $\phi_{\mathrm{c}}^{\prime}$ in (\ref{delta_prom}) is
inconsequential. The differentiation in the normalization factor $\left\vert
\partial E_{e}/\partial\phi_{\gamma}^{\prime}\right\vert ^{-1}$ should be done
using (\ref{E_e_prom}) and (\ref{angle_relation}), but before applying
(\ref{E_e_final}). This yields $E_{e}^{\prime}\partial E_{e}^{\prime}%
/\partial\phi_{\gamma}^{\prime}=p_{\gamma}^{\prime}p_{\gamma}\sin
\theta_{\gamma}^{\prime}\sin\theta_{\gamma}\sin\left(  \phi_{\gamma}^{\prime
}-\phi_{\gamma}\right)  $. Using (\ref{angle_relation}) again in order to
express $\sin\left(  \phi_{\gamma}^{\prime}-\phi_{\gamma}\right)  $ and then,
applying the energy conservation relation (\ref{E_e_final}), we obtain
$|\partial E_{e}^{\prime}/\partial\phi_{\gamma}^{\prime}|^{-1}=E_{e}^{\prime
}/(p_{\gamma}^{\prime}p_{\gamma}S^{1/2})$, where
\begin{align}
&  S\left(  \cos\Theta_{\gamma},\cos\theta_{\gamma}^{\prime},\cos
\theta_{\gamma}\right) \nonumber\\
&  =1-\cos^{2}\theta_{\gamma}-\cos^{2}\theta_{\gamma}^{\prime}-\cos^{2}%
\theta_{\gamma}+2\cos\Theta_{\gamma}\cos\theta_{\gamma}^{\prime}\cos
\theta_{\gamma} \label{S_symmetric}%
\end{align}
Implementing the entire procedure, we arrive at the integro-differential
equation (\ref{kinEq}), where $S$ is recast as (\ref{S}). This fully symmetric
function of three cosine arguments has a simple physical meaning. The
restriction\ $S\geq0$ covers all possible directions of the incident and
scattered photons characterized by the pairs of polar and axial angles
$\Theta_{\gamma},\Phi_{\gamma}$, $\theta_{\gamma},\phi_{\gamma}^{\prime}$,
$\theta_{\gamma},\phi_{\gamma}$, which are coupled by geometric relations like
(\ref{angle_relation}). The photon energy losses and the scattering angle
$\Theta_{\gamma}$ are rigidly coupled by (\ref{cos_Theta}). A photon with
given $\theta_{\gamma}^{\prime}$, $p_{\gamma}^{\prime}$ can be scattered into
another photon with given $\theta_{\gamma}$, $p_{\gamma}$, provided the photon
energies are within the allowed domains. This freedom is reached by a unique
choice of two symmetric values of $(\phi_{\gamma}-\phi_{\gamma}^{\prime})$.

\section{Solving kinetic equation (\ref{kinEq}%
)\label{Solving kinetic equation}}

\setcounter{equation}{0}

\subsection{General solution of multi-scattered
photons\label{General solution of multi-scattered photons}}

The LHS of kinetic equation~(\ref{kinEq}) is simple, but its RHS is rather
complicated, as described by (\ref{I_arr}). Our approach to solving the
kinetic equation is based on the physics of the process. The primary photons
with the monochromatic momentum distribution approximated by
(\ref{source_beam}) start efficiently scattering by air molecules within a
distance of $r\lesssim l=c/\nu$ from the source. For photon energies between
$0.3$ and $1.3\
%TCIMACRO{\unit{MeV}}%
%BeginExpansion
\operatorname{MeV}%
%EndExpansion
$, the maximum K-N differential cross-section is reached near the forward
scattering, $\Theta_{\gamma}=0$ and gradually decreases to the backscatter,
$\Theta_{\gamma}=\pi$, as shown in Fig.\ \ref{Fig:KN_figures}. This means that
photons scatter mainly through large angles, rather than broaden their
delta-function angular distribution around the primary beam. This suggests
considering the primary photon beam as a separate group that continues keeping
its delta-function distribution over long distances from the radiation source.
For the primary photons, we can neglect the first (\textquotedblleft
arrival\textquotedblright) term in the RHS of (\ref{kinEq}) but must keep the
last (\textquotedblleft departure\textquotedblright) term. Using the boundary
condition (\ref{source_beam}), we readily obtain for the primary photons the
exponentially decaying solution,
\begin{equation}
\left(  f_{\gamma}^{+}\right)  _{\mathrm{prim}}=\frac{I_{\gamma}^{(0)}%
\delta\left(  \cos\theta_{\gamma}-1\right)  }{8\pi^{2}cr^{2}}\sum_{s=1}%
^{k}\frac{e^{-r/l_{s}}\delta\left(  p_{\gamma}-p_{s}\right)  }{p_{s}^{2}%
}=\frac{I_{\gamma}^{(0)}\delta\left(  \rho\right)  }{8\pi^{2}p_{0}^{2}\rho
}\sum_{s=1}^{k}e^{-r/l_{s}}\delta\left(  p_{\gamma}-p_{s}\right)  .
\label{F_prim}%
\end{equation}
This solution gives the entire local source for scattered photons. The
solution for all propagating photons, including the primary beam, can be
sought as an infinite series%
\begin{equation}
f^{\pm}=f_{0}^{+}+f_{1}^{\pm}+f_{2}^{\pm}+f_{3}^{\pm}+\ldots=\sum
_{n=0}^{\infty}f_{n}^{\pm}, \label{f=}%
\end{equation}
where $f_{0}^{+}=\left(  f_{\gamma}^{+}\right)  _{\mathrm{prim}}$ and the
subscript $n$ denotes photons scattered from the primary beam $n$ times. For
each individual group of $n$-scattered photons, we obtain a recursive kinetic
equation where the source for scattering is provided by the $\left(
n-1\right)  $-th group,
\begin{equation}
\pm c\left(  1-\frac{\rho^{2}}{r^{2}}\right)  ^{1/2}\frac{\partial f_{n}^{\pm
}}{\partial r}+\nu f_{n}=\hat{L}_{r}f_{n-1}^{\prime\pm}, \label{recurr}%
\end{equation}
where $\hat{L}_{r}$ is the linear integral operator defined by (\ref{I_arr}).
The infinite series given by (\ref{f=}) converges. If $\nu$ includes no
absorption then Eq.~(\ref{recurr}) provides automatically the constancy of the
total photon flux through any closed surface surrounding the radiation source.
Using the boundary and matching conditions for the outgoing ($+$)\ and ingoing
($-$) particles described above, we obtain the exact recursive solution for
$f_{n}^{\pm}$ in terms of $f_{n-1}^{\pm}$, $l\equiv c/\nu$,%

\begin{align}
f_{n}^{-}  &  =\frac{\exp\left[  \left(  r^{2}-\rho^{2}\right)  ^{1/2}%
/l\right]  }{c}\sum_{\pm}\int_{r}^{\infty}\frac{\exp\left[  -\left(
y^{2}-\rho^{2}\right)  ^{1/2}/l\right]  \hat{L}_{y}f_{n-1}^{\prime\pm}%
}{\left(  y^{2}-\rho^{2}\right)  ^{1/2}}\ ydy,\nonumber\\
f_{n}^{+}  &  =\frac{\exp\left[  -\left(  r^{2}-\rho^{2}\right)
^{1/2}/l\right]  }{c}\sum_{\pm}\int_{\rho}^{\infty}\frac{\exp\left[  -\left(
y^{2}-\rho^{2}\right)  ^{1/2}/l\right]  \hat{L}_{y}f_{n-1}^{\prime\pm}%
}{\left(  y^{2}-\rho^{2}\right)  ^{1/2}}\ ydy\label{recur_relation}\\
&  +\frac{\exp\left[  -\left(  r^{2}-\rho^{2}\right)  ^{1/2}/l\right]  }%
{c}\sum_{\pm}\int_{\rho}^{r}\frac{\exp\left[  \left(  y^{2}-\rho^{2}\right)
^{1/2}/l\right]  \hat{L}_{y}f_{n-1}^{\prime\pm}}{\left(  y^{2}-\rho
^{2}\right)  ^{1/2}}\ ydy.\nonumber
\end{align}
This solution holds for $\rho\leq r$, while for $\rho>r$ all $f_{n}^{\pm}=0$.
As mentioned above, the integrations for $\hat{L}_{y}$ in (\ref{I_arr}) can be
performed it terms of $\left(  \cos\theta_{\gamma}^{\prime}\right)  $, but
before inserting the results to (\ref{recur_relation}) they have to be
transformed into functions of $\rho$ using (\ref{rho}). In the integrands\ of
(\ref{recur_relation}), the distance $r$ is replaced by an internal radial
variable $y$ in order to avoid a confusion with $r$ outside the integral. The
exponential factors describe $\left(  n-1\right)  $-photon losses due to their
scattering into the $n$-group with $|(r^{2}-\rho^{2})^{1/2}\pm(y^{2}-\rho
^{2})^{1/2}|$ describing the distance between two points on the $r$- and
$y$-shells along a straight-line photon trajectory corresponding to given
$\rho$ (the signs $\pm$ depend on whether the two points are on the same or on
different sides with respect to the trajectory median $r=\rho$, as shown in
Fig.\ \ref{Fig:Trajectories}). The summations over the ingoing and outgoing
particles in (\ref{recur_relation}) take into account the fact that photons
can be scattered through arbitrary scatter angles, so that any ingoing or
outgoing particle can be scattered into each of these groups.

The explicit solution for scattered photons with arbitrarily large $n$ can be
found by consecutively applying recursive relation (\ref{recur_relation})
starting from $n=1$ with $f_{0}^{\prime\pm}=\left(  f_{\gamma}^{+}\right)
_{\mathrm{prim}}$ given by Eq.~(\ref{F_prim}). At any given distance $r$,
photons with sufficiently large $n$ are produced mainly by ingoing photons
that originate from the primary source beam at much larger distances and only
accidentally come back. One might expect that they will come back less
frequently as $n$ increases, so that the infinite series (\ref{f=}) will
converge. Additionally, as $n$ increases, the photon energy distribution
gradually moves to lower energies. This effective "cooling" of scattered
photons will eventually put high-$n$ scattered photons into the keV range
where the photoemission absorption of photons starts dominating.

We can easily estimate the energy spread for $n$ times scattered photons. The
maximum energy of scattered photons, $p_{\gamma}=p_{\gamma}^{\prime}-I$, where
$I$ is the ionization potential, $I\sim15~\mathrm{eV}\lll p_{s}$, is reached
where $\theta_{\gamma}$ is close to $0^{\circ}$, but not too close \cite[Sect.
2.6.2.1]{Carron}. As a result, after $n$ scatterings of a primary $\gamma
$-quanta with $p_{\gamma}=p_{s}$, the maximum scattered photon energy reduces
only weakly, $p_{\max}^{(n)}=p_{s}-nI$, while in the framework of the K-N
theory $p_{\max}^{(n)}=p_{s}$. On the other hand, the minimum energy of
scattered photons is reached for strict backscatter, $\theta_{\gamma
}=180^{\circ}$ \cite{Carron}. After $n$ scatterings, the minimum energy,
$p_{\min}^{(n)}$, decreases by the following rule,%
\begin{equation}
p_{\min}^{(n)}=\frac{p_{\min}^{(n-1)}}{1+2p_{\min}^{(n-1)}/(m_{e}c)},
\label{rule}%
\end{equation}
starting from $p_{\min}^{(0)}=p_{0}$. Temporarily denoting $Y_{n}\equiv
m_{e}/(2p_{\min}^{(n)})$, we obtain from (\ref{rule}) $Y_{n}=1+Y_{n-1}$, so
that $Y_{n}=n+Y_{0}$, or%
\begin{equation}
p_{\min}^{(n)}=\frac{p_{s}}{1+2np_{s}/(m_{e}c)}=\frac{m_{e}c}{2n\left(
1+m_{e}c/(2np_{s})\right)  }. \label{p_min^n}%
\end{equation}
The energy of $n$-times scattered photons spreads between $p_{\min}^{(n)}$ and
$p_{\max}^{(n)}$. This spread gradually increases with $n$ due to lowering
$p_{\min}^{(n)}$. For $p_{s}\simeq1\
%TCIMACRO{\unit{MeV}}%
%BeginExpansion
\operatorname{MeV}%
%EndExpansion
$, $2p_{s}/(m_{e}c)\simeq4$, so that after 2--3 scatterings the minimum photon
energy $p_{\min}^{(n)}$ becomes almost inversely proportional to $n$,
$p_{\min}^{(n)}\simeq m_{e}c/(2n)\approx0.255%
%TCIMACRO{\unit{MeV}}%
%BeginExpansion
\operatorname{MeV}%
%EndExpansion
/n$. Hence, after $\sim10$ scatterings, the minimum energy falls well into the
range where photoemission absorption starts dominating. This low-energy photon
absorption creates an effective sink for multiply scattered photons.

\subsection{Solution for $n=1$\label{solution for n=1}}

Here we consider the lowest-order scattered photons, $n=1$. In the relatively
near zone, $r\lesssim l(p_{\gamma})$, the cutoff of the infinite chain by the
first term of the series represents a good approximation for the entire photon distribution.

For $n=1$, we obtain from (\ref{recur_relation}),
\begin{align}
f_{1}^{-}  &  =\frac{\exp\left[  \left(  r^{2}-\rho^{2}\right)  ^{1/2}%
/l\right]  }{c}\int_{r}^{\infty}\frac{\exp\left[  -\left(  y^{2}-\rho
^{2}\right)  ^{1/2}/l\right]  \left(  I_{\mathrm{arr}}\right)  _{1}}{\left(
y^{2}-\rho^{2}\right)  ^{1/2}}\ ydy,\nonumber\\
f_{1}^{+}  &  =\frac{\exp\left[  -\left(  r^{2}-\rho^{2}\right)
^{1/2}/l\right]  }{c}\int_{\rho}^{\infty}\frac{\exp\left[  -\left(  y^{2}%
-\rho^{2}\right)  ^{1/2}/l\right]  \left(  I_{\mathrm{arr}}\right)  _{1}%
}{\left(  y^{2}-\rho^{2}\right)  ^{1/2}}\ ydy\label{f_1}\\
&  +\frac{\exp\left[  -\left(  r^{2}-\rho^{2}\right)  ^{1/2}/l\right]  }%
{c}\int_{\rho}^{r}\frac{\exp\left[  \left(  y^{2}-\rho^{2}\right)
^{1/2}/l\right]  \left(  I_{\mathrm{arr}}\right)  _{1}}{\left(  y^{2}-\rho
^{2}\right)  ^{1/2}}\ ydy.\nonumber
\end{align}
Here $\left(  I_{\mathrm{arr}}\right)  _{1}$ is given by%
\begin{align}
\left(  I_{\mathrm{arr}}\right)  _{1}  &  =\frac{2Z_{\mathrm{air}}n_{m}%
r_{0}^{2}m_{e}c^{2}}{p_{\gamma}^{2}}\int_{p_{\gamma}}^{p_{\max}}\left(
\frac{p_{\gamma}}{p_{\gamma}^{\prime}}+\frac{p_{\gamma}^{\prime}}{p_{\gamma}%
}-\sin^{2}\Theta_{\gamma}\right)  dp_{\gamma}^{\prime}\nonumber\\
&  \times\int_{C^{-}}^{C^{+}}\frac{\left(  f_{\gamma}^{+\prime}\right)
_{\mathrm{prim}}d\left(  \cos\theta_{\gamma}^{\prime}\right)  }{\sqrt{S\left(
\cos\Theta_{\gamma},\cos\theta_{\gamma},\cos\theta_{\gamma}^{\prime}\right)
}}, \label{promezh}%
\end{align}
where $\cos\Theta_{\gamma}\left(  p_{\gamma},p_{\gamma}^{\prime}\right)  $ is
given by (\ref{cos_Theta}) and, accordingly,%
\begin{equation}
\sin^{2}\Theta_{\gamma}=L(p_{\gamma},p_{s})\equiv\left(  \frac{m_{e}%
c}{p_{\gamma}}-\frac{m_{e}c}{p_{s}}\right)  \left(  \frac{m_{e}c}{p_{s}}%
-\frac{m_{e}c}{p_{\gamma}}+2\right)  \label{sin^2Theta}%
\end{equation}%
\begin{equation}
\cos^{2}\Theta_{\gamma}=1-L(p_{\gamma},p_{s}) \label{cos^2}%
\end{equation}
In accord with (\ref{F_prim}), we obtain%
\begin{equation}
\left(  f_{\gamma}^{+\prime}\right)  _{\mathrm{prim}}=\frac{I_{\gamma}%
^{(0)}\delta\left(  \cos\theta_{\gamma}^{\prime}-1\right)  }{8\pi^{2}cr^{2}%
}\sum_{s=1}^{k}\frac{e^{-r/l_{s}}\delta\left(  p_{\gamma}^{\prime}%
-p_{s}\right)  }{p_{s}^{2}}. \label{f_prim'}%
\end{equation}
Unlike general equation\ (\ref{recur_relation}), (\ref{f_1}) contains no
summations over $\pm$ because the primary photon beams include no ingoing
photons. Integration in (\ref{promezh}) over $\cos\theta_{\gamma}^{\prime}$
should be done with care because the straightforward elimination of $\delta
$($\cos\theta_{\gamma}^{\prime}-1$) with $S\left(  \cos\Theta_{\gamma}%
,\cos\theta_{\gamma},\cos\theta_{\gamma}^{\prime}=1\right)  $ leads to a
singularity. To avoid it, we apply the following. Regardless of the specific
values of $\cos\Theta_{\gamma}$ and $\cos\theta_{\gamma}$, the integral
$\int_{C^{-}}^{C^{+}}(1/\sqrt{S})d\left(  \cos\theta_{\gamma}^{\prime}\right)
$ equals $\pi$, so that we obtain%
\begin{equation}
\lim_{\theta_{\gamma}^{\prime}\rightarrow0}\frac{1}{\sqrt{S\left(  \cos
\Theta_{\gamma},\cos\theta_{\gamma}^{\prime},\cos\theta_{\gamma}\right)  }%
}=\pi\delta\left(  \cos\Theta_{\gamma}-\cos\theta_{\gamma}\right)  .
\label{=pi_delta()}%
\end{equation}
and hence
\begin{subequations}
\label{I_arr_1}%
\begin{align}
\left(  I_{\mathrm{arr}}\right)  _{1}  &  =\frac{2Z_{\mathrm{air}}I_{\gamma
}^{(0)}n_{m}r_{0}^{2}m_{e}c}{8\pi p_{\gamma}^{2}\rho^{2}}\sum_{s=1}^{k}%
\frac{r_{s}e^{-r_{s}/l_{s}}}{p_{s}^{2}}\left(  1-\frac{\rho^{2}}{r_{s}^{2}%
}\right)  ^{1/2}\label{I_arr_1_fin_1}\\
&  \times\left(  \frac{p_{\gamma}}{p_{\gamma}^{\prime}}+\frac{p_{\gamma
}^{\prime}}{p_{\gamma}}+L(p_{\gamma},p_{s})\right)  \delta\left(
r-r_{s}\right)  ,\nonumber
\end{align}
where $l_{s}\equiv l(p_{s})=c/\nu(p_{s})$,
\end{subequations}
\begin{equation}
r_{s}\left(  \rho,p_{\gamma},p_{s}\right)  =\frac{\rho}{L^{1/2}}=\rho\left(
\frac{m_{e}c}{p_{\gamma}}-\frac{m_{e}c}{p_{s}}\right)  ^{-1/2}\left(
\frac{m_{e}c}{p_{s}}-\frac{m_{e}c}{p_{\gamma}}+2\right)  ^{-1/2}. \label{r_s}%
\end{equation}
These expressions have a simple physical meaning. Photons from the radiated
monochromatic and infinitely narrow beams are scattered into angles rigidly
coupled to the energy of the scattered photon. Given both the energy and
$\rho$ of the scattered photon, one can find the distance where such act of
scattering took place, $r=r_{s}$. The delta-function $\delta\left(
r-r_{s}\right)  $ demonstrates the uniqueness of this scattering distance. The
factor $e^{-r_{c}/l_{s}}$ describes the total attenuation of the primary beam
$s$ on its way from the radiation source to $r_{s}$.

Now we insert the expression for $\left(  I_{\mathrm{arr}}\right)  _{1}$ into
(\ref{f_1}), integrate with eliminating $\delta\left(  y-r_{s}\right)  $ and
use relation (\ref{r_s}): $y\rightarrow r_{s}$, $\left(  y^{2}-\rho
^{2}\right)  ^{1/2}\rightarrow\left(  r_{s}^{2}-\rho^{2}\right)  ^{1/2}$. The
result depends upon whether the photon momentum $p_{\gamma}$ is less or
greater than a critical value, $p_{\mathrm{cr}s}$, corresponding to the
$90^{\circ}$-scattering ($\cos\Theta_{\gamma}=0$) from the primary beam $s$.
For
\begin{equation}
p_{\mathrm{cr}s}\equiv\frac{p_{s}}{1+p_{s}/(m_{e}c)}\leq p_{\gamma}<p_{s}
\label{p_cr<p_gamma}%
\end{equation}
all singly-scattered particles are outgoing particles caused by forward
scattering ($\Theta_{\gamma}\leq90^{\circ}$), while for $p_{\gamma
}<p_{\mathrm{cr}s}$ scattered particles are either ingoing or outgoing
particles caused by backscatter ($90^{\circ}<\Theta_{\gamma}\leq180^{\circ}$).

Depending on whether $r>r_{s}$ or $r<r_{s}$, we will separate between `outer'
and `inner' photons. For $p_{\gamma}\geq p_{\mathrm{cr}s}$, all scattered
photons are outer ($r\geq r_{s}$) outgoing photons,%
\begin{equation}
\rho\leq L^{1/2}r=\left(  \frac{m_{e}c}{p_{\gamma}}-\frac{m_{e}c}{p_{s}%
}\right)  ^{1/2}\left(  \frac{m_{e}c}{p_{s}}-\frac{m_{e}c}{p_{\gamma}%
}+2\right)  ^{1/2}r, \label{rho<}%
\end{equation}
originating from only one scattering location with given $r_{s}$. We have then%
\begin{equation}
f_{1\mathrm{ext}}^{+}=\frac{2Z_{\mathrm{air}}I_{\gamma}^{(0)}n_{m}r_{0}%
^{2}m_{e}}{8\pi p_{\gamma}^{2}\rho^{2}}\sum_{s=1}^{k}\left(  \frac{p_{\gamma}%
}{p_{s}}+\frac{p_{s}}{p_{\gamma}}-L(p_{\gamma},p_{s})\right)  \frac
{A_{\mathrm{ext}}^{+}r_{s}}{p_{s}^{2}}, \label{external_p>p_gamma}%
\end{equation}
where the attenuation factor is%
\begin{equation}
A_{\mathrm{ext}}^{+}\left(  r,p_{\gamma},\rho\right)  =\exp\left\{  [\left(
r_{s}^{2}-\rho^{2}\right)  ^{1/2}-\left(  r^{2}-\rho^{2}\right)
^{1/2}]/l\left(  p_{\gamma}\right)  -r_{s}/l_{s}\right\}  \label{A_ext+}%
\end{equation}

For $p_{\gamma}<p_{\mathrm{cr}s}$, or more precisely, for
\begin{equation}
p_{\min}=\frac{p_{s}}{1+2p_{s}/(m_{e}c)}<p_{\gamma}<p_{\mathrm{cr}s}%
=\frac{p_{s}}{1+p_{s}/(m_{e}c)} \label{p_min<p_gamma<p_cr}%
\end{equation}
there are both outer outgoing and inner ($r<r_{s}$) photons,%
\begin{equation}
L^{1/2}r<\rho\leq r, \label{rho>}%
\end{equation}
either outgoing or ingoing from two different locations of given $r_{s}$. For
the outer photons defined by (\ref{rho<}), we have the same solution as
(\ref{external_p>p_gamma}). For the inner photons, we have two solutions for
the either ingoing or outgoing photons,%
\begin{equation}
f_{1\mathrm{int}}^{\pm}=\frac{2Z_{\mathrm{air}}I_{\gamma}^{(0)}n_{m}r_{0}%
^{2}m_{e}}{8\pi p_{\gamma}^{2}\rho^{2}}\sum_{s=1}^{k}\left(  \frac{p_{\gamma}%
}{p_{s}}+\frac{p_{s}}{p_{\gamma}}-L(p_{\gamma},p_{s})\right)  \frac
{A_{\mathrm{int}}^{\pm}r_{s}}{p_{s}^{2}}, \label{internal_p<p_gamma}%
\end{equation}
which differ by the attenuation factors only,%
\begin{equation}
A_{\mathrm{int}}^{\pm}\left(  r,p_{\gamma},\rho\right)  =\exp\left[  -\left(
r_{s}^{2}-\rho^{2}\right)  ^{1/2}/l\left(  p_{\gamma}\right)  \mp\left(
r^{2}-\rho^{2}\right)  ^{1/2}/l\left(  p_{\gamma}\right)  -r_{s}/l_{s}\right]
\label{A_int+-}%
\end{equation}
There are different signs in front of $\nu\left(  p_{\gamma}\right)  \left(
r^{2}-\rho^{2}\right)  ^{1/2}/c$ due to the fact that for the outgoing photons
the two points corresponding to $r$ and $r_{s}$ lie on the two opposite sides
from $r=\rho$, so that the two distances $\left(  r^{2}-\rho^{2}\right)
^{1/2}$ and $\left(  r_{s}^{2}-\rho^{2}\right)  ^{1/2}$ add, rather than
subtract, as they do for the ingoing photons.

Note that the only $r$-dependence in the above distributions is due to the
attenuation factors. As functions of the angular variable $\rho=L^{1/2}r$,
$f_{1}^{\pm}$ contains only a singular factor $r_{s}/\rho^{2}\propto1/\rho$.
Since all integrals for calculating the photon densities and fluxes contain
$d\left(  \cos\theta_{\gamma}\right)  \propto\rho d\rho$ this singularity is integrable.

For $n>1$, the formal general solution (\ref{f=}) with the recurrence
relations (\ref{recur_relation}) involves an increasing number of nested
integrations and becomes complicated. The solution for $n=1$, given by
equations (\ref{external_p>p_gamma})--(\ref{A_int+-}), will be the basis of
our analysis of the photon propagation in the near-source zone.

\end{document}